\documentclass[final]{elsarticle}
\usepackage{graphicx}
\newcommand{\beq}{\begin{equation}}
\newcommand{\beqa}{\begin{eqnarray}}
\newcommand{\eeq}{\end{equation}}
\newcommand{\eeqa}{\end{eqnarray}}
\newcommand{\abs}[1]{\vert#1\vert}
\newcommand{\bs}{{\;\;}}
\newcommand{\bl}{{\;\;\;\;}}

\renewcommand{\d}{{\rm d}}
\newcommand{\dpar}{\partial}
\newcommand{\ds}[1]{\displaystyle#1}
\newcommand{\e}{{\rm e}}
\newcommand{\eff}{{\rm eff}}
\newcommand{\eps}{\varepsilon}
\newcommand{\epsn}{{\eps_1\dots\eps_n}}
\newcommand{\euler}{{\bf C}}
\newcommand{\frad}[2]{\displaystyle{\displaystyle#1\over\displaystyle#2}}
\newcommand{\ii}{{\rm i}}
\renewcommand{\max}{{\rm max}}
\renewcommand{\min}{{\rm min}}
\newcommand{\meaneps}[1]{\langle#1\rangle_\epsn}
\renewcommand{\o}{{\omega}}
\newcommand{\ot}{\leftarrow}
\newcommand{\s}{{\sigma}}
\renewcommand{\sp}[1]{{{\hskip 2.4pt}#1{\hskip 2.4pt}}}
\newcommand{\tot}{\leftrightarrow}
\newcommand{\w}{\widehat}
\newcommand{\vareps}{{\rm var}_\epsn\,}
\newcommand{\C}{{\cal C}}
\newcommand{\N}{{\cal N}}

\begin{document}

\begin{frontmatter}

\title{On the frequencies of patterns of rises and falls}

\author{J. M. Luck}

\address{Institut de Physique Th\'eorique, URA 2306 of CNRS,
CEA Saclay, 91191~Gif-sur-Yvette~cedex, France}

\ead{jean-marc.luck@cea.fr}

\begin{abstract}
We investigate the probability of observing a given pattern of $n$
rises and falls in a random stationary data series.
The data are modelled
as a sequence of $n+1$ independent and identically distributed random numbers.
This probabilistic approach has a combinatorial equivalent,
where the data are modelled by a random permutation on $n+1$ objects.
The probability of observing a long pattern of rises and falls
decays exponentially with its length $n$ in general.
The associated decay rate $\alpha$
is interpreted as the embedding entropy of the pattern.
This rate is evaluated exactly for all periodic patterns.
In the most general case,
it is expressed in terms of a determinant
of generalized hyperbolic or trigonometric functions.
Alternating patterns have the smallest rate $\alpha_{{\rm min}}=\ln(\pi/2)=0.451582\dots$,
while other examples lead to arbitrarily large rates.
The probabilities of observing uniformly chosen random patterns
are demonstrated to obey multifractal statistics.
The typical value $\alpha_0=0.806361\dots$ of the rate
plays the role of a Lyapunov exponent.
A wide range of examples of patterns, either deterministic or random,
is also investigated.
\end{abstract}


\end{frontmatter}

\section{Introduction}

Consider a data series,
such as e.g.~the daily temperature at a given weather station over one year.
The most obvious features of such a data series are its rises and falls.
Physics and other branches of science provide plenty of examples of datasets
where the statistics of geometrical features,
such as maxima and minima, or rises and falls, is of central interest.
One example from statistical physics is provided by energy landscapes,
which are ubiquitously present in theoretical studies
of systems ranging from glasses to proteins~\cite{wales}.

In this work we investigate
the probability of observing a given pattern of~$n$
rises and falls in a random stationary data series.
This question has hardly been addressed so far in the physics literature,
in strong contrast with the statistics of extreme values,
which has recently attracted a lot of attention in many areas,
including random walks, disordered systems, growth processes
and random matrices~\cite{evs1,evs2,evs3,evs4,evs5,evs6}.

The setting of the present work is meant to provide a null model,
to which real data could be compared.
A first attempt has been made recently in this direction,
with the analysis of microarray time series data in genetics~\cite{FWB}.
We model the data series as a sequence
of $n+1$ i.i.d.~(independent and identically distributed) random numbers~$x_i$
drawn from a continuous distribution.
As these random numbers will only occur in inequalities,
their distribution can be chosen to be uniform on the unit interval.
This probabilistic approach is exposed in Section~\ref{proba}.
An equivalent combinatorial approach (see Section~\ref{combi})
is obtained by coarse-graining the random numbers
according to the permutation which brings them to an increasing order.
We are thus led to model the data as a uniformly chosen random permutation
on $n+1$ objects.
This line of thought dates back to the pioneering study of alternating permutations
by Andr\'e~\cite{DA1,DA2},
and it has since then been addressed regularly in the mathematical
literature~\cite{MM,ENT,N,DB,C,CS,A,HOF,GV,MA,MS,RBR,GS,BHR,BFW,MR,NBP,MAR}
(this list of references is not meant to be exhaustive).
To close, let us mention that the combinatorial approach to our problem
pertains to the more general topic of patterns in permutations,
which has been for long an active area
of discrete mathematics~\cite{bona,patterns1,patterns2}.

\section{Summary of results}

Our goal is to provide a comprehensive and self-contained exposition of the calculation
of the frequencies of patterns of rises and falls in a random stationary data series.
We aim at using a language accessible to a broad readership in statistical physics.
Let us give the detailed setup of this paper and summarize our findings.

The probabilistic and combinatorial approaches,
respectively exposed in Sections~\ref{proba} and~\ref{combi},
provide two equivalent definitions of
the probability $P_n(\epsn)$ of observing a given pattern $\epsn$ of $n$ rises and falls.
The equivalence between both approaches has already been underlined
in several works~\cite{FWB,MS,GS,BHR,BFW,NBP,MAR}.
It will become clear in the following that each approach has its advantages:
the probabilistic one is more suitable for analytical investigations,
while the combinatorial one results in a simple recursive structure,
lending itself to exact numerical calculations.

In Section~\ref{heuristic}
we show explicit results for small patterns (up to $n=4$).
We then present a heuristic analysis
demonstrating that the probability of observing a pattern
is essentially determined by its excursion,
as long as its length is modest.

In the remainder of the paper,
the emphasis is on asymptotic properties in the regime of most interest,
at least from the viewpoint of statistical physics,
i.e., where the length~$n$ of the pattern is large.
In this regime the probability $P_n(\epsn)$ typically falls off exponentially as
\beq
P_n\sim\e^{-\alpha n}.
\label{rate}
\eeq
The decay rate $\alpha$ will be our central object of interest.
This quantity can be viewed as the {\it embedding entropy}
of the binary pattern $\epsn$, i.e., the entropic cost per unit length
for embedding this pattern into a sequence of i.i.d.~random numbers.
It is worth noticing that the above definition is entirely parameter-free.
If all the $2^n$ patterns of length $n$ had equal probabilities $P_n=2^{-n}$,
the rate would be constant and equal to $\alpha=\ln 2$.
The observed wide range of possible values of the rate $\alpha$,
from $\alpha_\min=\ln(\pi/2)=0.451582\dots$ to infinity,
testifies the richness of the problem.

Periodic patterns are investigated in
sections~\ref{alt},~\ref{palt}, and~\ref{parb}.
As already mentioned, the subject is an old classic of discrete mathematics.
Our comprehensive approach allows us to recover many known results
by more elementary means,
and often to express them in simpler terms.
Section~\ref{alt} is a self-contained presentation of some of the beauties
of the historical example of alternating patterns,
for which the rate $\alpha$ assumes its minimal value~$\alpha_\min$.
Section~\ref{palt} deals with the family of $p$-alternating periodic patterns,
whose motif (unit cell) consists of $p-1$ rises followed by a fall.
The rate reads $\alpha=\ln z_0$,
where $z_0$ is the smallest real positive zero
of a generalized trigonometric function.
In Section~\ref{parb} we show how the rate~$\alpha$ can be evaluated exactly
for an arbitrary periodic pattern, with any period $p\ge2$:
$z_0$ is now the smallest zero of
a determinant of generalized hyperbolic or trigonometric functions,
whose size is at most $p/2$.
Many examples are treated explicitly.

The rest of the paper covers entirely novel areas.
Sections~\ref{aperiodic} and~\ref{chirping} serve as an intermezzo.
In Section~\ref{aperiodic} we deal with examples of aperiodic patterns
which are built from three classical self-similar sequences:
Fibonacci, Thue-Morse, and Rudin-Shapiro.
The probabilities $P_n$ exhibit an exponential decay,
characterized by a well-defined rate $\alpha$,
modulated by a fractal amplitude which reflects the self-similarity
of the underlying sequence.
Section~\ref{chirping} is devoted to chirping patterns,
consisting mostly of rises, whereas falls are more and more scarce (or vice versa).
In this case the probabilities $P_n$ are found to decay super-exponentially.
Their asymptotic form is predicted more precisely
in the situation of most interest where the density of falls follows a power law.

Section~\ref{random} is devoted to the heart of the problem,
namely the statistics of the probabilities $P_n$
if patterns are chosen in various ensembles of random patterns of fixed length $n$.
The uniform ensemble,
where all patterns are considered with equal weights,
is studied thoroughly.
The probabilities $P_n$ of generic patterns
have the typical rate $\alpha_0=0.806361\dots$
The latter number can be interpreted as a Lyapunov exponent.
The whole set of probabilities $P_n$ is shown to obey multifractal statistics,
with a non-trivial spectrum of multifractal dimensions $f(\alpha)$,
increasing from $f(\alpha_\min)=0$ to $f(\alpha_0)=1$.
Other ensembles of random patterns of fixed length $n$,
namely the ensemble at fixed concentration $c$ of rises
and a symmetric Markovian ensemble defined by a persistence probability $r$,
are also investigated.
The probabilities $P_n$ now generically decay according to effective typical rates
$\beta(c)$ and $\gamma(r)$, which depend continuously on the ensemble parameters.

Two appendices are respectively devoted to
the explicit correspondence between the combinatorial
and probabilistic approaches~(\ref{appa})
and to generalized hyperbolic and trigonometric functions~(\ref{appb}).

\section{Probabilistic approach}
\label{proba}

The probabilistic approach goes as follows.
The data series is modeled by a sequence of $n+1$ i.i.d.~random numbers
$x_i$ ($i=0,\dots,n$),
drawn from the uniform distribution on the unit interval $[0,\,1]$.
This sequence of random numbers
yields a pattern $\epsn$ of $n$ rises and falls defined as follows.
For $i=1,\dots,n$:
\beq
\matrix{
\mbox{If $x_i>x_{i-1}$, there is a rise at the $i$th place, and $\eps_i=+$,}
\hfill\cr
\mbox{If $x_i<x_{i-1}$, there is a fall at the $i$th place, and $\eps_i=-$.}
\hfill
}
\eeq

\begin{figure}[!ht]
\begin{center}
\includegraphics[angle=-90,width=.4\linewidth]{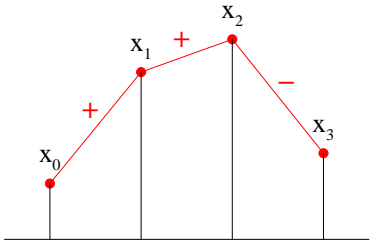}
\caption{\small
An example of a configuration of 4 random numbers.}
\label{example}
\end{center}
\end{figure}

Let us start with the example shown in Figure~\ref{example}.
This configuration obeys the inequalities $x_0<x_1<x_2>x_3$,
and therefore yields the pattern $++-$.
The probability of observing this pattern reads
\beqa
P_3(++-)
\!\!\!&=&\!\!\!
\int_0^1\d x_0
\int_{x_0}^1\d x_1
\int_{x_1}^1\d x_2
\int_0^{x_2}\d x_3
\nonumber\\
\!\!\!&=&\!\!\!
\int_0^1\d x_3
\int_{x_3}^1\d x_2
\int_0^{x_2}\d x_1
\int_0^{x_1}\d x_0
\nonumber\\
\!\!\!&=&\!\!\!
\frac{1}{8}.
\label{exap}
\eeqa

Let us now turn to the general case.
The probability of an arbitrary pattern $\epsn$
can be calculated recursively by conditioning on $x_n$.
Let $f_n(x)\,\d x$ be the probability that the sequence $x_0,\dots,x_n$
yields the pattern $\epsn$ and that $x<x_n<x+\d x$.
The conditional densities $f_n(x)$ obey the recursion relations:
\beq
\matrix{
\mbox{If $\eps_n=+$, then}
\bs{\ds f_n(x)=\int_0^x f_{n-1}(y)\,\d y},\hfill\cr\cr
\mbox{If $\eps_n=-$, then}
\bs{\ds f_n(x)=\int_x^1 f_{n-1}(y)\,\d y},\hfill
}
\label{prec}
\eeq
with the initial condition $f_0(x)=1$.
It follows that $f_n(x)$ is a polynomial in $x$ with degree $n$.
We have
\beq
P_n(\epsn)=\int_0^1 f_n(x)\,\d x.
\label{ppp}
\eeq
This recursive scheme has already been described in~\cite{GS,BHR,MAR}.
For the example of Figure~\ref{example}, we get
\beqa
f_1(x)
\!\!\!&=&\!\!\!
\int_0^x\d y=x,
\nonumber\\
f_2(x)
\!\!\!&=&\!\!\!
\int_0^x f_1(y)\,\d y=\frac{x^2}{2},
\nonumber\\
f_3(x)
\!\!\!&=&\!\!\!
\int_x^1f_2(y)\,\d y=\frac{1-x^3}{6},
\nonumber\\
P_3(++-)
\!\!\!&=&\!\!\!
\int_0^1 f_3(x)\,\d x=\frac{1}{8}.
\eeqa
The second of the nested integral expressions~(\ref{exap}) is thus recovered.

\section{Combinatorial approach}
\label{combi}

In the combinatorial approach
the data series is modeled by a permutation~$\s$,
chosen uniformly among the $(n+1)!$ permutations on $n+1$ objects
labelled $i=0,\dots,n$.
Such a permutation yields a pattern $\epsn$, defined as follows.
For $i=1,\dots,n$:
\beq
\matrix{
\mbox{If $\s_i>\s_{i-1}$, there is a rise at the $i$th place, and $\eps_i=+$,}
\hfill\cr
\mbox{If $\s_i<\s_{i-1}$, there is a fall at the $i$th place, and $\eps_i=-$.}
\hfill
}
\eeq
The pattern $\epsn$ is said to be the {\it up-down signature}
of the permutation $\s$.

Let us again consider the example shown in Figure~\ref{example}.
We have
\beq
x_0<x_3<x_1<x_2.
\label{ineqs}
\eeq
Let us coarse-grain this configuration of random numbers
by representing it as a permutation $\s$.
The rule is that the inverse permutation $\s^{-1}$
gives the order of indices in the inequalities~(\ref{ineqs}).
We thus have $\s^{-1}=(0312)$,\footnote{We use the one-line notation
$\s=(\s_0\s_1\dots\s_n)$.}
and so $\s=(0231)$,
which indeed yields the pattern $++-$.

How many permutations on 4 objects yield the pattern $++-$~?
This simple question can be solved by inspection.
There are 3 such permutations: (0132), (0231) and (1230).
We thus recover the result (see~(\ref{exap}))
\beq
P_3(++-)=\frac{3}{4!}=\frac{1}{8}.
\label{exac}
\eeq
It is indeed clear that all the $4!$ permutations on 4 objects
are equally probable.
This observation demonstrates (on an example)
the equivalence between the probabilistic approach
of Section~\ref{proba} and the present combinatorial one
(see~\cite{FWB,MS,GS,BHR,BFW,NBP,MAR}).

Let us now turn to the general case.
The number $A_n(\epsn)$ of permutations on $n+1$ objects yielding
a given pattern $\epsn$ of $n$ rises and falls
can be calculated by an efficient recursive scheme,
which seems to have been discovered several times independently~\cite{DB,GV,MA}.
The gist of the method is similar to that of the probabilistic approach.
It consists in conditioning the permutation~$\s$ on its last entry $\s_n$,
and to relate the patterns formed by $n+1$ and $n$ objects.
Let $a_{n,j}$ be the number of permutations which yield
the pattern $\epsn$ and have $\s_n=j$.
These numbers can be shown to obey the recursion relations
\beq
\matrix{
\mbox{If $\eps_n=+$, then}
\left\{\matrix{a_{n,0}=0,\hfill\cr
a_{n,j}=a_{n,j-1}+a_{n-1,j-1}\quad(j=\overrightarrow{1,\dots,n}),}\right.\hfill\cr\cr
\mbox{If $\eps_n=-$, then}
\left\{\matrix{a_{n,n}=0,\hfill\cr
a_{n,j}=a_{n,j+1}+a_{n-1,j}\quad(j=\overleftarrow{0,\dots,n-1}),}\right.\hfill
}
\label{arec}
\eeq
with the initial condition $a_{0,0}=1$.
Arrows indicate the order in which the recursion relations have to be used.
We thus build a triangular array of integers:
\beq
\matrix{
a_{0,0}\cr
a_{1,0}\tot a_{1,1}\cr
a_{2,0}\tot a_{2,1}\tot a_{2,2}\cr
a_{3,0}\tot a_{3,1}\tot a_{3,2}\tot a_{3,3}\cr
a_{4,0}\tot a_{4,1}\tot a_{4,2}\tot a_{4,3}\tot a_{4,4}
}
\label{atri}
\eeq
At the $n$th row, all arrows go from left to right if $\eps_n=+$,
and from right to left if $\eps_n=-$ (see~(\ref{bous}) for an example).

The requested numbers of permutations
\beq
A_n(\epsn)=\sum_{j=0}^n a_{n,j}
\label{asum}
\eeq
can also be read off from the array $a_{n,j}$.
We have indeed
\beq
\matrix{
\mbox{If $\eps_{n+1}=+$, then}\bs A_n=a_{n+1,n+1},\hfill\cr
\mbox{If $\eps_{n+1}=-$, then}\bs A_n=a_{n+1,0}.\hfill
}
\label{arecsum}
\eeq
Finally, the probability that a random permutation
on $n+1$ objects yields the pattern $\epsn$ reads
\beq
P_n(\epsn)=\frac{A_n(\epsn)}{(n+1)!}.
\label{ppc}
\eeq

To close, let us write down explicitly
the correspondence between the probabilistic and combinatorial approaches.
For a given pattern $\epsn$,
the probabilistic approach involves the $n$th degree polynomial $f_n(x)$,
which has $n+1$ coefficients,
while the combinatorial one involves the $n+1$ integers $a_{n,j}$.
These two sets of numbers carry the same information.
The precise correspondence, to be established in~\ref{appa},
is summarized by the formula
\beq
f_n(x)=\sum_{j=0}^n a_{n,j}\,\frac{x^j(1-x)^{n-j}}{j!(n-j)!}.
\label{pa}
\eeq

\section{Explicit results for small patterns}
\label{heuristic}

Table~\ref{small} presents explicit results
for patterns of length up to $n=4$.
For each length~$n$, the $2^n$ patterns are listed in lexicographical order.
For each pattern $\epsn$,
the Table gives the number $A_n(\epsn)$ of permutations yielding that pattern
and the corresponding probability $P_n(\epsn)$,
obtained by means of~(\ref{arec}),~(\ref{asum}), and~(\ref{ppc}).
These numbers exhibit the expected up-down and left-right symmetries.
They also manifest a scatter which increases rapidly with~$n$.

\begin{table}[!ht]
\begin{center}

$n=1$
\smallskip

\begin{tabular}{|c|c|c||c|c|c|}
\hline
$\eps_1$ & $A_1$ & $P_1$ &
$\eps_1$ & $A_1$ & $P_1$\\
\hline
$+$ & 1 & $1/2$ & $-$ & 1 & $1/2$\\
\hline
\end{tabular}

\medskip
$n=2$
\smallskip

\begin{tabular}{|c|c|c||c|c|c|}
\hline
$\eps_1\eps_2$ & $A_2$ & $P_2$ &
$\eps_1\eps_2$ & $A_2$ & $P_2$\\
\hline
$++$ & 1 & $1/6$ &
$-+$ & 2 & $1/3$\\
$+-$ & 2 & $1/3$ &
$--$ & 1 & $1/6$\\
\hline
\end{tabular}

\medskip
$n=3$
\smallskip

\begin{tabular}{|c|c|c||c|c|c|}
\hline
$\eps_1\eps_2\eps_3$ & $A_3$ & $P_3$ &
$\eps_1\eps_2\eps_3$ & $A_3$ & $P_3$\\
\hline
$+++$ & 1 & $1/24$	& $-++$ & 3 & $1/8$\\
$++-$ & 3 & $1/8$	& $-+-$ & 5 & $5/24$\\
$+-+$ & 5 & $5/24$	& $--+$ & 3 & $1/8$\\
$+--$ & 3 & $1/8$	& $---$ & 1 & $1/24$\\
\hline
\end{tabular}

\medskip
$n=4$
\smallskip

\begin{tabular}{|c|c|c||c|c|c|}
\hline
$\eps_1\eps_2\eps_3\eps_4$ & $A_4$ & $P_4$ &
$\eps_1\eps_2\eps_3\eps_4$ & $A_4$ & $P_4$\\
\hline
$++++$ & 1 & $1/120$	& $-+++$ & 4 & $1/30$\\
$+++-$ & 4 & $1/30$	& $-++-$ & 11 & $11/120$\\
$++-+$ & 9 & $3/40$	& $-+-+$ & 16 & $2/15$\\
$++--$ & 6 & $1/20$	& $-+--$ & 9 & $3/40$\\
$+-++$ & 9 & $3/40$	& $--++$ & 6 & $1/20$\\
$+-+-$ & 16 & $2/15$	& $--+-$ & 9 & $3/40$\\
$+--+$ & 11 & $11/120$	& $---+$ & 4 & $1/30$\\
$+---$ & 4 & $1/30$	& $----$ & 1 & $1/120$\\
\hline
\end{tabular}

\end{center}
\caption{Explicit results
for all patterns of rises and falls of length up to $n=4$:
numbers $A_n(\epsn)$ of permutations and probabilities $P_n(\epsn)$.}
\label{small}
\end{table}

The data shown in Table~\ref{small} can be analyzed in the following heuristic way.
The probability $P_n(\epsn)$ can be expected to be strongly penalized
for patterns which make large excursions in the vertical direction.
This notion can be formalized as follows.
To the pattern $\epsn$ we associate a random walk with steps $\eps_i$, i.e.,
\beq
h_i=h_{i-1}+\eps_i,
\eeq
with $h_0=0$,
and we define the {\it excursion} of the pattern
as the variance of the position of that walk:
\beqa
\Delta_n^2(\epsn)
\!\!\!&=&\!\!\!
\frac{1}{n+1}\sum_{i=0}^n h_i^2
-\left(\frac{1}{n+1}\sum_{i=0}^n h_i\right)^2
\nonumber\\
\!\!\!&=&\!\!\!
\frac{n(n+2)}{6(n+1)}+\frac{2}{(n+1)^2}\sum_{1\le i<j\le n}i(n+1-j)\eps_i\eps_j.
\eeqa

For a fixed length $n$,
the patterns with the largest excursion are the two straight ones,
consisting only of rises, or of falls, for which we have
\beq
\Delta_n^2=\frac{n(n+2)}{12}.
\eeq
These patterns also have the smallest probabilities.
Consider for definiteness the rising pattern ($\epsn=+\cdots+$).
The probabilistic approach yields
\beq
f_n(x)=\frac{x^n}{n!},
\eeq
while the combinatorial approach yields
\beq
a_{n,j}=\delta_{n,j},\qquad A_n=1,
\eeq
where the Kronecker symbol $\delta_{n,j}$ equals 1 if $j=n$ and 0 else.
There is indeed one single permutation yielding the rising pattern,
namely the identity ($\s_j=j$ for all $j$).
Both approaches consistently give
\beq
P_n=\frac{1}{(n+1)!}.
\label{rising}
\eeq
This factorial decay corresponds to a logarithmically divergent rate
\beq
\alpha(n)\approx\ln n-1,
\label{alog}
\eeq
and so $\alpha_\max=\infty$.

The patterns with the smallest excursion are the two alternating ones,
$+-+-\cdots$ and $-+-+\cdots$, for which we have
\beq
\Delta_n^2=\frac{1}{4}\quad(n\bs\hbox{odd}),\qquad
\Delta_n^2=\frac{n(n+2)}{4(n+1)^2}\quad(n\bs\hbox{even}).
\eeq
These patterns, which will be investigated thoroughly in Section~\ref{alt},
are also known to have the largest probabilities~\cite{N,GV}.
The corresponding rate $\alpha$ takes its minimum value
$\alpha_\min=\ln(\pi/2)$ (see~(\ref{a+-})).

\begin{figure}[!ht]
\begin{center}
\includegraphics[angle=-90,width=.5\linewidth]{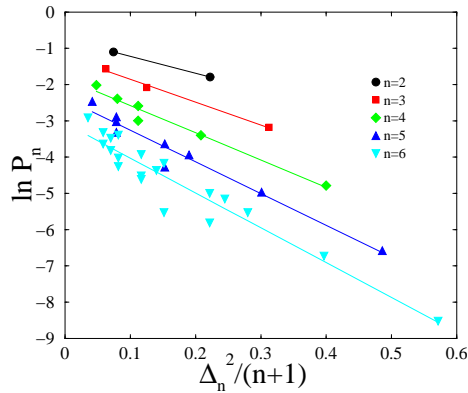}
\caption{\small
Logarithmic plot of probability $P_n$
against reduced excursion $\Delta_n^2/(n+1)$
for all patterns of length $n=2$ to~6.
Symbols: data for individual patterns.
Straight lines: least-square fits.}
\label{heu}
\end{center}
\end{figure}

More generally, we observe (see Figure~\ref{heu})
that the probability $P_n$ and the ex\-cur\-sion~$\Delta_n^2$
are strongly anti-correlated.

\begin{figure}[!ht]
\begin{center}
\includegraphics[angle=-90,width=.5\linewidth]{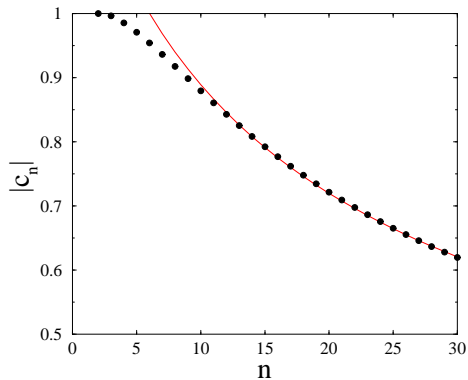}
\caption{\small
Absolute correlation coefficient $\abs{c_n}$ between $\Delta_n^2$ and $\ln P_n$
against length $n$.
Black symbols: exact numerical data for lengths up to $n=30$.
Full red line: fit of the data for $n\ge10$ to the form $c_n=a(n+n_0)^{-1/2}$.}
\label{cor}
\end{center}
\end{figure}

The observed negative correlation however progressively fades away
as the length of patterns is increased.
Figure~\ref{cor} shows a plot
of exact numerical data\footnote{Throughout the following,
the expression {\it exact numerical data} refers to results
obtained by iterating numerically the combinatorial recursion~(\ref{arec}).}
for the absolute correlation coefficient $\abs{c_n}$ between $\Delta_n^2$ and $\ln P_n$
for lengths up to $n=30$.
The fit to the data suggests that this coefficient falls off asymptotically as $n^{-1/2}$.

A statistical analysis of the probabilities $P_n$
adapted to their behavior for large lengths $n$,
based on the multifractal formalism,
will be presented in Section~\ref{uni}.

\section{Alternating patterns}
\label{alt}

This Section is devoted to the historical example of alternating patterns $+-+-+-\cdots$,
studied long ago by Andr\'e in the framework of alternating permutations~\cite{DA1,DA2}.

\subsection{Probabilistic approach}

We have $\eps_{2k}=-$ and $\eps_{2k+1}=+$,
and so the recursion~(\ref{prec}) for the polynomials $f_n(x)$ takes the form
\beq
f_{2k+1}(x)=\int_0^xf_{2k}(y)\,\d y,\qquad
f_{2k}(x)=\int_x^1f_{2k-1}(y)\,\d y.
\eeq
We thus obtain
\beqa
&&f_0(x)=1,\bl
f_1(x)=x,\bl
f_2(x)=\frac{1}{2}(1-x^2),\bl
\nonumber\\
&&f_3(x)=\frac{x}{6}(3-x^2),\bl
f_4(x)=\frac{1}{24}(1-x^2)(5-x^2),\bs\dots
\eeqa
Integrating over $x$, using~(\ref{ppp}) and~(\ref{ppc}),
we recover the first few entries of the sequence~(\ref{A000111}).

\subsection{Combinatorial approach}

The construction~(\ref{atri}) of the array of integers $a_{n,j}$
takes the following form:
\beq
\matrix{
\sp{0}\cr
\sp{0}\to\sp{1}\cr
\sp{1}\ot\sp{1}\ot\sp{0}\cr
\sp{0}\to\sp{1}\to\sp{2}\to\sp{2}\cr
\sp{5}\ot\sp{5}\ot\sp{4}\ot\sp{2}\ot\sp{0}\cr
\sp{0}\to\sp{5}\to10\to14\to16\to16\cr
61\ot61\ot56\ot46\ot32\ot16\ot\sp{0}
}
\label{bous}
\eeq
The above boustrophedon construction~\cite{MSY} seems to date back to Seidel~\cite{sei}.
The word {\it boustrophedon},
from two words of ancient Greek meaning {\it ox} and {\it turn},
applies to writing systems
where the direction of drawing is alternatively changed line after line
from right to left and left to right, somewhat like the motion of a plough in a field.
The resulting sequence for the numbers of alternating permutations:
\beqa
&&A_0=1,\bl
A_1=1,\bl
A_2=2,\bl
A_3=5,\bl
A_4=16,\bl
A_5=61,\bl
\nonumber\\
&&A_6=272,\bl
A_7=1385,\bl
A_8=7936,\bl
A_9=50521,\bs\dots
\label{A000111}
\eeqa
to be investigated below,
is referred to as sequence A000111 in the OEIS~\cite{OEIS},
where many further references and links are given.
The integers $A_n$ have been referred to as the up-down Euler-Bernoulli numbers
or the Entringer numbers,
and they have received many combinatorial interpretations.

\subsection{Generating-series method}
\label{gsm}

The probabilistic recursion~(\ref{prec})
can be solved analytically by means of generating series,
for any periodic pattern.
We explain the method in detail in the present case of alternating patterns,
and apply it to other patterns in Sections~\ref{palt} and~\ref{parb}.
Let us mention that such generating series can also be derived
by purely combinatorial means.
That route however requires an advanced knowledge
of the theory of symmetric functions~\cite{MR}.

Introduce the generating series
\beq
F(z,x)=\sum_{n\ge0}f_n(x)z^n
\eeq
and
\beq
\Pi(z)=\sum_{n\ge0}P_nz^n=\sum_{n\ge0}A_n\frac{z^n}{(n+1)!}=\int_0^1F(z,x)\,\d x.
\eeq
In the present case, it is advisable to set
\beq
F(z,x)=F_0(z,x)+F_1(z,x),
\eeq
where
\beq
F_0(z,x)=\sum_{k\ge0}f_{2k}(x)z^{2k},\qquad
F_1(z,x)=\sum_{k\ge0}f_{2k+1}(x)z^{2k+1}.
\eeq
The recursion~(\ref{prec}) translates to the coupled integral equations
\beq
F_0(z,x)=1+z\int_x^1F_1(z,y)\,\d y,\quad
F_1(z,x)=z\int_0^xF_0(z,y)\,\d y,
\label{cint}
\eeq
or, equivalently, to the coupled differential equations
\beq
\frac{\dpar F_0}{\dpar x}=-zF_1,\qquad
\frac{\dpar F_1}{\dpar x}=zF_0,
\eeq
with boundary conditions $F_0(z,1)=1$, $F_1(z,0)=0$,
and, finally, to the single differential equation
\beq
\frac{\dpar^2 F_0}{\dpar x^2}=-z^2F_0,
\eeq
with boundary conditions $F_0(z,1)=1$, $\dpar F_0(z,0)/\dpar x=0$.
The solution to the latter equation reads
\beq
F_0(z,x)=\frac{\cos zx}{\cos z},\qquad
F_1(z,x)=\frac{\sin zx}{\cos z}.
\eeq
Using~(\ref{cint}), we have
\beq
\Pi(z)=\frac{1}{z}\left(F_1(z,1)+F_0(z,0)-1\right).
\eeq
We thus obtain the explicit expression\footnote{$\tan z=\sin z/\cos z$,
$\sec z=1/\cos z$.}
\beq
\Pi(z)=\frac{\sin z+1-\cos z}{z\cos z}=\frac{1}{z}\left(\tan z+\sec z-1\right).
\label{pialt}
\eeq
Splitting the above result, we get
\beqa
\tan z
\!\!\!&=&\!\!\!
\sum_{k\ge0}P_{2k}z^{2k+1}=\sum_{k\ge0}A_{2k}\frac{z^{2k+1}}{(2k+1)!},
\nonumber\\
\sec z
\!\!\!&=&\!\!\!
1+\sum_{k\ge0}P_{2k+1}z^{2k+2}=1+\sum_{k\ge0}A_{2k+1}\frac{z^{2k+2}}{(2k+2)!}.
\label{andre}
\eeqa
This is the main result obtained by Andr\'e~\cite{DA1,DA2}.

To be complete, let us mention that the above expressions imply the following relationships
between the probabilities $P_n$ of alternating patterns,
or equivalently the Euler-Bernoulli or Entringer numbers $A_n$ of alternating permutations,
and the Bernoulli numbers $B_n$, the Euler numbers $E_n$,
and the value of Riemann's zeta function at even integers:
\beqa
P_{2k}
\!\!\!&=&\!\!\!
(-)^k\frac{2^{2k+2}(2^{2k+2}-1)}{(2k+2)!}\,B_{2k+2}
=\frac{2(2^{2k+2}-1)}{\pi^{2k+2}}\,\zeta(2k+2),
\nonumber\\
P_{2k+1}
\!\!\!&=&\!\!\!
(-)^{k+1}\frac{E_{2k+2}}{(2k+2)!}.
\eeqa

The asymptotic behavior of the probabilities $P_n$ is governed by the first pole
at $\pi/2$ of the generating series $\Pi(z)$.
We thus obtain the exponential decay
\beq
P_n\approx\frac{8}{\pi^2}\left(\frac{2}{\pi}\right)^n.
\label{resalt}
\eeq
The rate $\alpha$ assumes its minimal value
\beq
\alpha_\min=\ln(\pi/2)=0.451582\dots
\label{a+-}
\eeq
This result already appears in the physics literature, albeit in some disguised form,
in an investigation of the one-dimensional Ising spin glass at zero temperature
by Derrida and Gardner~\cite{DG}.

\section{$p$-alternating patterns}
\label{palt}

We pursue our investigation with the periodic patterns with any period $p\ge2$
whose motif (unit cell) consists of $p-1$ rises followed by one single fall.
In other words, we have $\eps_{kp+q}=+$ for $q=1,\dots,p-1$ and $\eps_{kp}=-$.
These $p$-alternating patterns have been investigated in~\cite{C,CS,BHR,MR,MAR}.

\subsection{General solution}

The generating-series method of Section~\ref{gsm} extends
to $p$-alternating patterns as follows.
Setting
\beq
F(z,x)=\sum_{q=0}^{p-1}F_q(z,x),\qquad
F_q(z,x)=\sum_{k\ge0}f_{kp+q}(x)z^{kp+q},
\label{setting}
\eeq
the recursion~(\ref{prec}) translates to the coupled integral equations
\beqa
&&F_0(z,x)=1+z\int_x^1F_1(z,y)\,\d y,
\nonumber\\
&&F_q(z,x)=z\int_0^xF_{q-1}(z,y)\,\d y\quad(q=1,\dots,p-1),
\eeqa
or, equivalently, to the differential equation
\beq
\frac{\dpar^p F_0}{\dpar x^p}=-z^pF_0,
\eeq
with boundary conditions $F_0(z,1)=1$ and $\dpar^q F_0(z,0)/\dpar x^q=0$ $(q=1,\dots,p-1)$.
The solution reads
\beq
F_q(z,x)=\frac{T_{p,q}(zx)}{T_{p,0}(z)}\quad(q=0,\dots,p-1),
\eeq
where the $T_{p,q}$ are the generalized trigonometric functions introduced in~\ref{appb}.
The generating series $\Pi(z)$ of the probabilities $P_n$ then reads
\beq
\Pi(z)=\frac{1}{zT_{p,0}(z)}\left(\sum_{q=1}^{p-1}T_{p,q}(z)+1-T_{p,0}(z)\right).
\label{pip}
\eeq
This result has been obtained by purely combinatorial means
by Mendes and Remmel~\cite{MR}.
The corresponding numbers $A_n$ of $p$-alternating partitions
have been called generalized Euler numbers
and investigated in~\cite{LM1,LM2}.

For $p\ge3$, the expression~(\ref{pip}) has two main novel features
with respect to its analogue~(\ref{pialt}) for alternating permutations ($p=2$).
These properties also apply to arbitrary periodic patterns,
to be investigated in Section~\ref{parb}.

\begin{itemize}

\item[(i)]
The denominator $T_{p,0}(z)$ is an entire function of $z^p$.
It therefore has $p$ smallest zeros sitting at the vertices of a regular $p$-gon,
i.e., $z_q=z_0\o^q$ $(q=0,\dots,p-1)$,
where~$z_0$ is real positive, and $\o=\e^{2\pi\ii/p}$ (see~(\ref{odef})).
We thus obtain an asymptotic decay of the probabilities $P_n$ of the form
\beq
P_n\approx\C_n\,\e^{-\alpha n},
\label{pc}
\eeq
where the rate $\alpha$ is given by
\beq
\alpha=\ln z_0,
\label{resal}
\eeq
while the other zeros $z_q$ are responsible for the occurrence
of a periodic amplitude $\C_n$ of $n$ with period $p$
(to be illustrated in Figure~\ref{period4} below).
This periodic modulation does not occur in the case of alternating patterns ($p=2$).
The expression~(\ref{pialt}) indeed has no pole at $z=-z_0=-\pi/2$,
as its numerator also vanishes there.

\item[(ii)]
If periodic patterns are deduced one from the other by a cyclic permutation,
such as those defined by the motifs $(-++)$, $(+-+)$ and $(++-)$,
they share the same rate $\alpha$, but different periodic amplitudes.
This phenomenon too is absent for $p=2$, because of up-down symmetry.

\end{itemize}

\subsection{The case $p=3$ $(++-)$}

The generating series $\Pi(z)$ reads (see~(\ref{t3}))
\beq
\Pi(z)=\frad
{3-\e^{-z}-2\,\e^{z/2}\cos\frac{z\sqrt{3}}{2}+2\sqrt{3}\,\e^{z/2}\sin\frac{z\sqrt{3}}{2}}
{z\left(\e^{-z}+2\,\e^{z/2}\cos\frac{z\sqrt{3}}{2}\right)}.
\label{3alt}
\eeq
We have
\beq
z_0=1.849812\dots,\qquad\alpha=0.615084\dots
\label{a++-}
\eeq
The resulting sequence for the numbers of 3-alternating permutations:
\beqa
&&A_0=1,\bl
A_1=1,\bl
A_2=1,\bl
A_3=3,\bl
A_4=9,\bl
A_5=19,\bl
\nonumber\\
&&A_6=99,\bl
A_7=477,\bl
A_8=1513,\bl
A_9=11259,\bs\dots
\label{A178963}
\eeqa
is referred to as sequence A178963 in the OEIS~\cite{OEIS},
where an expression equivalent to~(\ref{3alt}) is also given.

\subsection{The case $p=4$ $(+++-)$}

The generating series $\Pi(z)$ reads (see~(\ref{t4}))
\beq
\Pi(z)=\frac
{\sinh\zeta\sin\zeta+\sqrt{2}\cosh\zeta\sin\zeta-\cosh\zeta\cos\zeta+1}
{z\cosh\zeta\cos\zeta},
\label{4alt}
\eeq
with $\zeta=z/\sqrt{2}$.
We have
\beq
z_0=\frac{\pi}{\sqrt{2}}=2.221441\dots,\qquad\alpha=0.798156\dots
\label{a+++-}
\eeq
The resulting sequence for the numbers of 4-alternating permutations:
\beqa
&&A_0=1,\bl
A_1=1,\bl
A_2=1,\bl
A_3=1,\bl
A_4=4,\bl
A_5=14,\bl
\nonumber\\
&&A_6=34,\bl
A_7=69,\bl
A_8=496,\bl
A_9=2896,\bs\dots
\label{A178964}
\eeqa
is referred to as sequence A178964 in the OEIS~\cite{OEIS},
where an expression equivalent to~(\ref{4alt}) is also given.

\begin{figure}[!ht]
\begin{center}
\includegraphics[angle=-90,width=.5\linewidth]{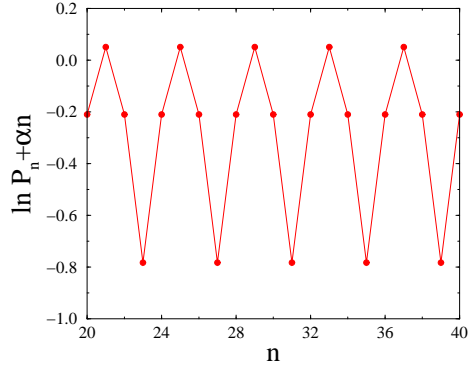}
\caption{\small
Plot of $\ln P_n+\alpha n\approx\ln\C_n$
for 4-alternating patterns against the pattern length $n$ in some range,
illustrating the modulation of the exponential decay
of the probabilities~$P_n$ by the periodic amplitude $\C_n$ (see~(\ref{pc})).}
\label{period4}
\end{center}
\end{figure}

Figure~\ref{period4} illustrates the modulation of the exponential decay
of the probabilities $P_n$ by a periodic amplitude $\C_n$ of the pattern length $n$
with period $p$ (see~(\ref{pc})).

\subsection{Behavior of the rate as a function of $p$}

It is of interest to investigate the behavior of the rate $\alpha$
as a function of the period $p$ of the $p$-alternating patterns.

\begin{figure}[!ht]
\begin{center}
\includegraphics[angle=-90,width=.5\linewidth]{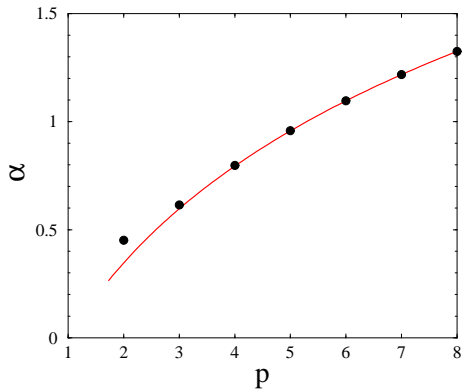}
\caption{\small
Rate $\alpha$ for $p$-alternating patterns against their period $p$.
Black symbols: exact values.
Full red line: result~(\ref{largep}) of the large-$p$ approximation.}
\label{alphap}
\end{center}
\end{figure}

In the regime where $p$ is large,
it is legitimate to approximate the full generalized trigonometric function $T_{p,0}$
by the first two terms of its series expansion~(\ref{tseries}),
i.e., $T_{p,0}(z)\approx1-z^p/p!$.
We thus obtain the estimates
\beq
z_0\approx(p!)^{1/p},\qquad
\alpha\approx\frac{\ln p!}{p},
\label{largep}
\eeq
implying the asymptotic logarithmic growth
\beq
\alpha\approx\ln p-1.
\label{alphalog}
\eeq
Figure~\ref{alphap} shows a plot of the rate $\alpha$ against $p$.
The large-$p$ approximation is observed to be extremely accurate,
except for $p=2$ and 3.
Corrections to~(\ref{largep}),
due to higher-order terms in the expansion~(\ref{tseries}),
are indeed exponentially small in $p$.

\section{Arbitrary periodic patterns}
\label{parb}

We now turn to the analysis of an arbitrary periodic pattern with period $p\ge2$,
whose motif consists of $p-\nu$ rises and $\nu$ falls.
If $\nu=0$ or $\nu=p$, the pattern is one of the straight ones
discussed in Section~\ref{heuristic}.
Without loss of generality,
we can assume $1\le\nu\le p/2$ (otherwise exchange the roles of rises and falls).
Finally, as we are chiefly interested in the rate $\alpha$,
we can use cyclic invariance
(see point (ii) below~(\ref{resal}))
to ensure that the motif ends with a fall.

\subsection{General form of the solution}
\label{form}

The generating-series method of Section~\ref{gsm} extends to the general case
as follows.
Using again the splitting~(\ref{setting}) of the generating series $F(z,x)$,
the recursion~(\ref{prec}) translates to the differential equation
\beq
\frac{\dpar^p F_0}{\dpar x^p}=(-1)^\nu z^pF_0,
\label{diff}
\eeq
with $F_0(z,1)=1$,
while the $p-1$ other boundary conditions ($q=1,\dots,p-1$) read:
\beq
\matrix{
\mbox{If $\eps_{p-q}=+$, then}\bs\dpar^q F_0/\dpar x^q(z,0)=0,\hfill\cr
\mbox{If $\eps_{p-q}=-$, then}\bs\dpar^q F_0/\dpar x^q(z,1)=0.\hfill
}
\eeq

The differential equation~(\ref{diff}) appears in~\cite{BHR} and~\cite{MAR}.
Let us show that its solution can be simply expressed in terms
of the generalized hyperbolic and trigonometric functions introduced in~\ref{appb}.
Consider for definiteness the case where $\nu$ is even.
We are led to look for $F_0(z,x)$ as a linear combination
of generalized hyperbolic functions:
\beq
F_0(z,x)=\sum_q C_q(z) H_{p,q}(zx),
\label{expan}
\eeq
where the sum runs over the $\nu$ indices $q$ such that $\eps_{p-q}=-$.
Imposing the boundary conditions at $x=1$ yields a system
of $\nu$ linear equations for the coefficients $C_q(z)$.
The key object is the denominator $\Delta(z)$ of the solution to the latter system.
It is a $\nu\times\nu$ determinant of generalized hyperbolic functions $H_{p,q}(z)$.
If $\nu$ is odd, the same holds with generalized trigonometric functions.
In both cases, $\Delta(z)$ is an entire function of $z^p$.
As a consequence, the asymptotic decay of the probabilities $P_n$
is given by the formulas~(\ref{pc}),~(\ref{resal}) in full generality,
where $z_0$ is the smallest real positive zero of $\Delta(z)$.

\subsection{An explicit example: the motif $(++--)$}

Let us illustrate the above formalism on the example of the motif $(++--)$.
We have $p=4$ and $\nu=2$.
The expansion~(\ref{expan}) reads
\beq
F_0(z,x)=C_0(z)H_{4,0}(zx)+C_1(z)H_{4,1}(zx).
\eeq
The boundary conditions at $x=1$ are
\beqa
&&C_0(z)H_{4,0}(z)+C_1(z)H_{4,1}(z)=1,
\nonumber\\
&&C_0(z)H_{4,3}(z)+C_1(z)H_{4,0}(z)=0.
\label{sys}
\eeqa
We thus obtain
\beq
C_0(z)=\frac{H_{4,0}(z)}{\Delta(z)},\qquad
C_1(z)=-\frac{H_{4,3}(z)}{\Delta(z)},
\eeq
where $\Delta(z)$ is the determinant of the $2\times2$ system~(\ref{sys}), i.e.,
\beq
\Delta(z)=H_{4,0}(z)^2-H_{4,1}(z)H_{4,3}(z)=\frac{1}{2}(\cosh z\cos z+1).
\eeq
We obtain after some algebra
\beqa
2z\Delta(z)\Pi(z)
\!\!\!&=&\!\!\!
(\sin z-\cos z+1)(\cosh z-1)
\nonumber\\
\!\!\!&+&\!\!\!
(\sin z+\cos z+1)\sinh z.
\label{s++--}
\eeqa
We have
\beq
z_0=1.875104\dots,\qquad\alpha=0.628664\dots
\label{a++--}
\eeq
The resulting sequence of numbers of permutations
\beqa
&&A_0=1,\bl
A_1=1,\bl
A_2=1,\bl
A_3=3,\bl
A_4=6,\bl
A_5=26,\bl
\nonumber\\
&&A_6=71,\bl
A_7=413,\bl
A_8=1456,\bl
A_9=10576,\bs\dots
\label{A058258}
\eeqa
is referred to as sequence A058258 in the OEIS~\cite{OEIS},
where an expression equivalent to~(\ref{s++--}) is also given.

The four examples~(\ref{A000111}),~(\ref{A178963}),~(\ref{A178964}),
and~(\ref{A058258})
seem to exhaust the list of sequences of numbers of permutations
with prescribed up-down signatures
which are given in the OEIS~\cite{OEIS},
together with the corresponding exponential generating series.
The sequence of numbers of 5-alternating permutations is also given there
as sequence A181936, but without the corresponding generating series.
The latter is given by~(\ref{pip}) for $p=5$.
This result is however not very useful from a merely computational viewpoint,
as the expressions of generalized hyperbolic and trigonometric functions
for periods $p>4$ are rather cumbersome.

\subsection{Two falls per motif}

Let us consider a general pattern with period $p$
whose motif contains two falls separated by distances $a$ and $b$.
We have therefore $\eps_q=+$ except $\eps_a=\eps_{a+b}=-$, and so $p=a+b$ and $\nu=2$.

The formalism of Section~\ref{form} leads to
\beq
\Delta(z)=\left\vert\matrix{H_{p,0}(z) & H_{p,a}(z)\cr H_{p,b}(z) & H_{p,0}(z)}\right\vert
=H_{p,0}(z)^2-H_{p,a}(z)H_{p,b}(z).
\label{two}
\eeq

It is again worth investigating the behavior of the rate $\alpha$
in the regime where $a$ and $b$ are large.
Following the lines which led to~(\ref{largep}),
we thus obtain the estimate
\beq
z_0\approx(a!\,b!)^{1/p}.
\label{largepr}
\eeq
We have thus again a logarithmic growth law, of the form
\beq
\alpha\approx\ln p_\eff-1,\qquad
p_\eff=\left(a^a\,b^b\right)^{1/p}.
\eeq
The effective period $p_\eff$ interpolates between $p_\eff\approx p$ if the falls
are close to each other (i.e., $a\ll p$ or $b\ll p$)
and $p_\eff\approx p/2$ if $a$ and $b$ are nearly equal (i.e., $a\approx b\approx p/2$).

\subsection{Three falls per motif}

Consider a pattern with period $p$
whose motif contains three falls separated by distances $a$, $b$ and~$c$.
We have therefore $p=a+b+c$ and $\nu=3$.
The formalism of Section~\ref{form} leads to
\beq
\Delta(z)=\left\vert\matrix{
T_{p,0}(z)\hfill & T_{p,c}(z)\hfill & T_{p,b+c}(z)\hfill\cr
-T_{p,a+b}(z)\hfill & T_{p,0}(z)\hfill& T_{p,b}(z)\hfill\cr
-T_{p,a}(z)\hfill & -T_{p,a+c}(z)\hfill& T_{p,0}(z)\hfill
}\right\vert.
\label{three}
\eeq
Notice that all the minus signs are located in the lower triangle.

\subsection{Four falls per motif}

Consider a pattern with period $p$
whose motif contains four falls separated by distances $a$, $b$, $c$ and $d$.
We have $p=a+b+c+d$ and $\nu=4$.
The formalism of Section~\ref{form} leads to
\beq
\Delta(z)=\left\vert\matrix{
H_{p,0}(z)\hfill & H_{p,d}(z)\hfill & H_{p,c+d}(z)\hfill & H_{p,b+c+d}(z)\hfill\cr
H_{p,a+b+c}(z)\hfill & H_{p,0}(z)\hfill & H_{p,c}(z)\hfill & H_{p,b+c}(z)\hfill\cr
H_{p,a+b}(z)\hfill & H_{p,a+b+d}(z)\hfill & H_{p,0}(z)\hfill & H_{p,b}(z)\hfill\cr
H_{p,a}(z)\hfill & H_{p,a+d}(z)\hfill & H_{p,a+c+d}(z)\hfill & H_{p,0}(z)\hfill\cr
}\right\vert.
\label{four}
\eeq

The general structure of our prediction
for the rate of arbitrary periodic patterns
clearly emerges from the expressions~(\ref{two}),~(\ref{three}) and~(\ref{four}).

\subsection{Summary}

Let us summarize our findings.
For an arbitrary periodic pattern, with any period $p\ge2$,
the probabilities $P_n$ have an asymptotic exponential decay,
modulated by a periodic function of the pattern length $n$ with period $p$
(see~(\ref{pc})).
The rate~$\alpha$ is given by~(\ref{resal})
in terms of the smallest real positive zero $z_0$
of a $\nu\times\nu$ determinant $\Delta(z)$
of generalized hyperbolic or trigonometric functions,
whose size $\nu$ is at most $p/2$.
The rate is invariant under cyclic permutations of the motif,
under the exchange of rises and falls (up-down symmetry)
and under reversal (left-right symmetry).
Finally, any two periodic patterns are related by one of the above symmetries,
they share the same rate, but different periodic modulations in general.

Table~\ref{pertable} gives a list of irreducible motifs up to period $p=7$,
with the exact numerical values of the corresponding rates.
For each period, all symmetries have been used
in order to identify a minimal set of patterns.
The motifs thus obtained have been ordered
according to increasing values of $\alpha$.
For all motifs with periods up to $p=4$,
where analytical results have been derived,
the Table also lists the number of the equation giving the generating series $\Pi(z)$
and the OEIS reference~\cite{OEIS} of the sequence~$A_n$ of numbers of permutations.

\begin{table}[!ht]
\begin{center}

$p=2$
\smallskip

\begin{tabular}{|c|c|c|c|}
\hline
motif & $\alpha$ & equation & OEIS \\
\hline
$(+-)$ & 0.451582 & (\ref{pialt}) & A000111 \\
\hline
\end{tabular}

\medskip
$p=3$
\smallskip

\begin{tabular}{|c|c|c|c|}
\hline
motif & $\alpha$ & equation & OEIS \\
\hline
$(++-)$ & 0.615084 & (\ref{3alt}) & A178963\\
\hline
\end{tabular}

\medskip
$p=4$
\smallskip

\begin{tabular}{|c|c|c|c|}
\hline
motif & $\alpha$ & equation & OEIS \\
\hline
$(++--)$ & 0.628664 & (\ref{s++--}) & A058258\\
$(+++-)$ & 0.798156 & (\ref{4alt}) & A178964\\
\hline
\end{tabular}

\medskip
$p=5$
\smallskip

\begin{tabular}{|c|c||c|c|}
\hline
motif & $\alpha$ &
motif & $\alpha$\\
\hline
$(++-+-)$ & 0.542722 &
$(++++-)$ & 0.958296\\
$(+++--)$ & 0.740839 &
&\\
\hline
\end{tabular}

\medskip
$p=6$
\smallskip

\begin{tabular}{|c|c||c|c|}
\hline
motif & $\alpha$ &
motif & $\alpha$\\
\hline
$(++-+--)$ & 0.581879 &
$(++++--)$ & 0.866884\\
$(+++-+-)$ & 0.669441 &
$(+++++-)$ & 1.096722\\
$(+++---)$ & 0.799654 &
&\\
\hline
\end{tabular}

\medskip
$p=7$
\smallskip

\begin{tabular}{|c|c||c|c|}
\hline
motif & $\alpha$ &
motif & $\alpha$\\
\hline
$(++-+-+-)$ & 0.516159 &
$(++++-+-)$ & 0.797400\\
$(++-++--)$ & 0.619535 &
$(++++---)$ & 0.889929\\
$(+++-+--)$ & 0.674316 &
$(+++++--)$ & 0.988659\\
$(+++-++-)$ & 0.718458 &
$(++++++-)$ & 1.217921\\
\hline
\end{tabular}

\end{center}
\caption{List of irreducible motifs of periodic patterns up to period $p=7$,
with exact numerical values of the rate $\alpha$
characterizing the asymptotic exponential decay of the probabilities $P_n$.
The equation number of the corresponding generating series $\Pi(z)$
and the OEIS reference~\cite{OEIS} of the sequence $A_n$ of numbers of permutations
are also given when applicable ($p\le4$).}
\label{pertable}
\end{table}

\section{Deterministic aperiodic patterns}
\label{aperiodic}

We have seen that the probabilities $P_n$ of periodic patterns
have an exponential decay, modulated by an oscillatory function of the pattern length $n$.
The same property extends to a much wider class of patterns.

This holds in particular for patterns which are built
from aperiodic sequences, such as e.g.~the Fibonacci sequence.
An interesting class of deterministic aperiodic sequences
are the self-similar sequences generated
by substitutions on a finite alphabet~\cite{que}.
These sequences exhibit an intermediate degree of order between periodic and random.
Investigations of the properties of various physical models
defined on such sequences are reviewed in~\cite{review}.

We shall successively consider three classic examples of such sequences:
Fibonacci, Thue-Morse, and Rudin-Shapiro
(see~\cite{que} for historical references and details).
In each case, we demonstrate by means of exact numerical calculations
that the probabilities~$P_n$ exhibit an exponential decay,
with a well-defined rate $\alpha$,
modulated by an aperiodic amplitude, either bounded or very slowly increasing,
whose fractal structure reflects the self-similarity of the underlying sequence
(see Figures~\ref{fibo},~\ref{morse}, and~\ref{rudin}).

\subsubsection*{Fibonacci sequence.}

It is generated by the substitution
\beq
S_{\rm Fib}:\;\left\{\matrix{A\to AB\cr B\to A\hfill}\right.
\eeq
on two letters.
Starting with $A$ and iterating the above rules,
we obtain the Fibonacci sequence
$ABAABABAABAAB\dots$
This sequence is quasiperiodic, and provides a one-dimensional analogue
of the icosahedral quasicrystals discovered in 1984~\cite{she}.

\begin{figure}[!ht]
\begin{center}
\includegraphics[angle=-90,width=.5\linewidth]{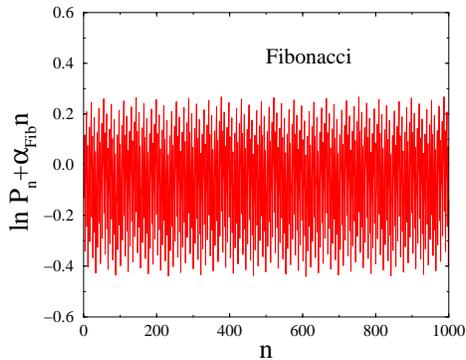}
\caption{\small
Plot of $\ln P_n+\alpha_{\rm Fib}n$ against length $n$ for Fibonacci patterns.}
\label{fibo}
\end{center}
\end{figure}

Interpreting every letter $A$ as a rise ($\eps=+$)
and every letter $B$ as a fall ($\eps=-$),
we have thus constructed a family of patterns of any length $n$.
Figure~\ref{fibo} shows a plot of exact numerical data
for the quantity $\ln P_n+\alpha_{\rm Fib}n$
for patterns with length up to 1000.
The following accurate value of the rate has been obtained by fitting the data:
\beq
\alpha_{\rm Fib}=0.562168\dots
\eeq

\subsubsection*{Thue-Morse sequence.}

\begin{figure}[!ht]
\begin{center}
\includegraphics[angle=-90,width=.5\linewidth]{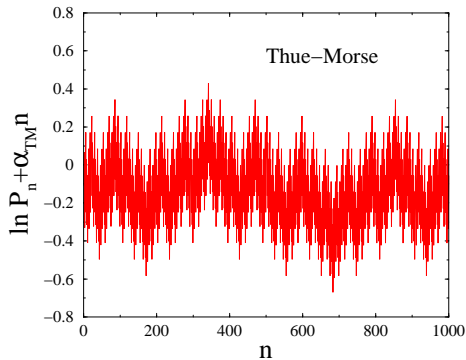}
\caption{\small
Plot of $\ln P_n+\alpha_{\rm TM}n$ against length $n$ for Thue-Morse patterns.}
\label{morse}
\end{center}
\end{figure}

It is generated by the substitution
\beq
S_{\rm TM}:\;\left\{\matrix{A\to AB\cr B\to BA\hfill}\right.
\eeq
which again acts on two letters.
Starting with $A$ and iterating the above rules,
we obtain the Thue-Morse sequence
$ABBABAABBAABABBA\dots$
This sequence has many specific properties,
including a purely singular continuous Fourier transform~\cite{que}.
Figure~\ref{morse} shows a plot of the quantity $\ln P_n+\alpha_{\rm TM}n$
for the patterns thus defined, with
\beq
\alpha_{\rm TM}=0.583018\dots
\eeq

\subsubsection*{Rudin-Shapiro sequence.}

\begin{figure}[!ht]
\begin{center}
\includegraphics[angle=-90,width=.5\linewidth]{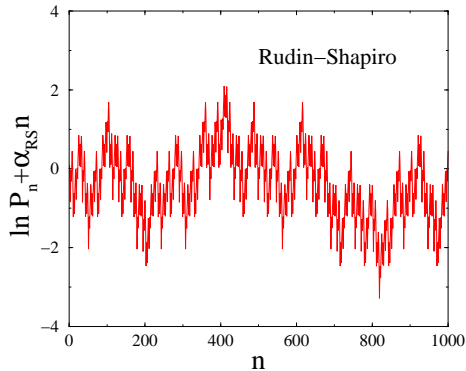}
\caption{\small
Plot of $\ln P_n+\alpha_{\rm RS}n$ against length $n$ for Rudin-Shapiro patterns.}
\label{rudin}
\end{center}
\end{figure}

It is generated by the substitution
\beq
S_{\rm RS}:\;\left\{\matrix{A\to AC\cr B\to DC\cr C\to AB\cr D\to DB\hfill}\right.
\eeq
acting on four letters.
Starting with $A$ and iterating the above rules,
we obtain the Rudin-Shapiro sequence
$ACABACDCACABDBAB\dots$
The rule is now to read every $A$ or $C$ as a rise ($\eps=+$)
and every $B$ or $D$ as a fall ($\eps=-$).
The binary sequence thus obtained again has many peculiar properties~\cite{que}.
Figure~\ref{rudin} shows a plot of the quantity $\ln P_n+\alpha_{\rm RS}n$
for the patterns thus defined, with
\beq
\alpha_{\rm RS}=0.780693\dots
\eeq

\section{Chirping patterns}
\label{chirping}

We have seen that the two straight patterns
exhibit a factorial decay of the probabilities $P_n$ (see~(\ref{rising})),
formally corresponding to an infinite decay rate.
More generally, patterns consisting mostly of rises,
whereas falls become more and more scarce (or vice versa),
can be expected to yield a super-exponential decay of the probabilities.
We refer to these patterns as chirping,
because the density of falls slowly goes to zero.
Besides being a birdsong,
a {\it chirp} is indeed also a signal whose frequency varies slowly in time.

Consider a chirping pattern consisting mostly of rises.
The position $n(k)$ of the $k$th fall grows faster than linearly in $k$,
and so the distance $p(k)$ between the $k$th fall and the $(k-1)$st one
grows indefinitely with $k$.
If $p$ were a large fixed number, the pattern would be periodic,
and so the probabilities $P_n$ would decay exponentially,
with a rate $\alpha\approx\ln p-1$ growing logarithmically with the period $p$
(see~(\ref{alphalog})).
Now, in the presence of a slowly varying `period' $p(k)$,
it seems natural to estimate the probabilities as
\beq
\ln P_n\approx-\sum_{n(k)<n}p(k)(\ln p(k)-1).
\label{periods}
\eeq
This estimate generalizes the result~(\ref{largepr}).
It also agrees with an exact upper bound for $P_n$,
which has been conjectured to be asymptotically exact
in the regime where all the distances $p(k)$ are large~\cite{BFW}.

The above prediction can be made more precise in the case of a power-law scaling
\beq
n(k)\approx C\,k^b,
\label{chpower}
\eeq
with a scaling exponent $b>1$,
so that the distance $p(k)$ diverges itself as a power law:
\beq
p(k)\approx bC\,k^{b-1}.
\eeq
Evaluating the sum in~(\ref{periods}) as an integral,
we thus obtain a super-exponential decay of the form
\beq
\ln P_n\approx-n\left(\frac{b-1}{b}\ln n+\ln b-2+\frac{1+\ln C}{b}\right).
\label{chres}
\eeq
The associated rate formally diverges logarithmically, as
\beq
\alpha(n)\approx\frac{b-1}{b}\ln n.
\label{acont}
\eeq
As the scaling exponent $b$ can take any value in the range $b>1$,
there is a continuum of logarithmically divergent effective rates $\alpha(n)$,
bounded by the worst case of the straight patterns (see~(\ref{alog})).

We have checked the above prediction
against exact numerical data in the following two cases.

\subsubsection*{Square chirping patterns.}

Falls occur at places given by the squares of the integers:
$n=k^2$ ($k=1,2,\dots$).
We have $b=2$, $C=1$, and so~(\ref{chres}) reads
\beq
\ln P_n\approx-\frac{n}{2}\,(\ln(4n)-3).
\label{chsquare}
\eeq

\subsubsection*{Triangular chirping patterns.}

Falls occur at places given by the triangular numbers:
$n=k(k+1)/2$ ($k=1,2,\dots$).
We have $b=2$, $C=1/2$, and so~(\ref{chres})~reads
\beq
\ln P_n\approx-\frac{n}{2}\,(\ln(2n)-3).
\label{chtri}
\eeq

\begin{figure}[!ht]
\begin{center}
\includegraphics[angle=-90,width=.5\linewidth]{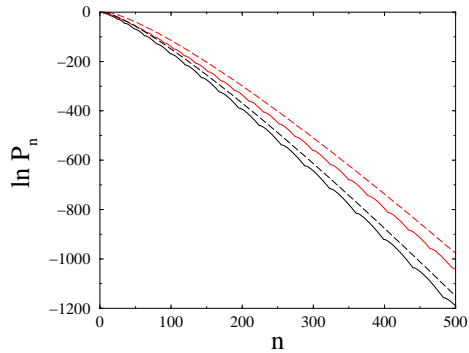}
\caption{\small
Logarithmic plot of the probabilities $P_n$ of chirping patterns
against length~$n$.
Full undulating lines: exact data.
Lower (black): square patterns.
Upper (red): triangular patterns.
Dashed lines: asymptotic results~(\ref{chsquare}),~(\ref{chtri}).}
\label{chirps}
\end{center}
\end{figure}

Figure~\ref{chirps} shows a logarithmic plot of the probabilities $P_n$ in both cases.
The data exhibit a super-exponential decay which is correctly described
by the asymptotic results~(\ref{chsquare}),~(\ref{chtri}) (dashed lines),
together with an undulation induced by the distribution of falls.

\section{Random patterns}
\label{random}

\subsection{Uniform patterns: multifractal properties}
\label{uni}

We now turn to the statistical analysis of the probabilities $P_n$
of observing random patterns chosen in various ensembles.
In this Section we consider a pattern chosen uniformly among
the $2^n$ patterns $\epsn$ of fixed length $n$.

\begin{figure}[!ht]
\begin{center}
\includegraphics[angle=-90,width=.5\linewidth]{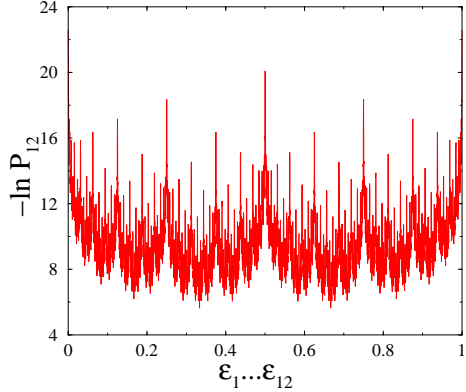}
\caption{\small
Plot of $-\ln P_{12}\approx12\,\alpha$
of all patterns of length $n=12$ in lexicographical order.}
\label{pplot}
\end{center}
\end{figure}

Figure~\ref{pplot} shows a plot of $-\ln P_{12}$
for the 4096 patterns of length $n=12$, listed in lexicographical order.
This plot gives a picture of the behavior of the rate~$\alpha$
as a function of the pattern.
We have indeed $-\ln P_{12}\approx12\,\alpha$.
The rate~$\alpha$ is observed to exhibit a very erratic behavior,
with structures at all scales.
This suggests that {\it multifractal analysis}~\cite{multif1,multif2,multif3}
may provide the appropriate framework for a quantitative analysis of these data.

Let us recall the basics of the multifractal formalism
in a framework adapted to the present case.
The key object is the partition function
\beq
Z_n(q)=2^n\meaneps{P_n^q}=\sum_\epsn P_n(\epsn)^q,
\eeq
where we have introduced the notation
\beq
\meaneps{X}=\frac{1}{2^n}\sum_\epsn X(\epsn).
\eeq
The real parameter $q$ plays the role of the inverse temperature $\beta$,
while the role of the energy of the configuration $\epsn$
is played by $-\ln P_n(\epsn)$.

If $q=2,3,\dots$ is a positive integer,
$Z_n(q)$ can be interpreted as the probability that~$q$ random permutations,
chosen independently and uniformly,
have the same up-down signature, i.e., yield the same pattern~\cite{MS}.

The set $\{P_n\}$ is said to be multifractal if the partition function
obeys an exponential law of the form
\beq
Z_n(q)\sim 2^{-n\tau(q)},
\label{Zsca}
\eeq
at least in some range of values of $q$.
The function $\tau(q)$ is the analogue of a free energy.
The normalization of the probabilities implies $Z_n(1)=1$,
hence the obvious result
\beq
\meaneps{P_n}=2^{-n},
\label{pave}
\eeq
and $\tau(1)=0$.
We set
\beq
\tau(q)=(q-1)D_q,
\eeq
where $D_q$ is the generalized (R\'enyi) dimension of order $q$.
If all the $2^n$ patterns of length $n$ had equal probabilities $P_n=2^{-n}$,
we would have $Z_n(q)=2^{-n(q-1)}$, and therefore $D_q=1$ for all $q$.

The scaling law~(\ref{Zsca}) is commonly interpreted
in terms of a multifractal spectrum of rates $\alpha$.
For a fixed $\alpha$, consider the set $\N(\alpha,\d\alpha)$
of patterns $\epsn$ such that $n\alpha<-\ln P_n(\epsn)<n(\alpha+\d\alpha)$.
In a generic multifractal situation,
the set $\N(\alpha,\d\alpha)$ has a well-defined dimension $f(\alpha)$,
meaning that its size (the number of its elements) grows exponentially as
\beq
\abs{\N(\alpha,\d\alpha)}\sim2^{nf(\alpha)}.
\eeq
The partition function may thus be estimated as
\beq
Z_n(q)\sim\int_0^\infty\e^{-nq\alpha}\,2^{nf(\alpha)}\,\d\alpha.
\eeq
Evaluating the integral by the saddle-point method,
we obtain the property
that the functions $\tau(q)$ and $f(\alpha)$ are related to each other
by a Legendre transform:\footnote{Here and in the following, primes denote derivatives.}
\beq
\tau(q)+f(\alpha)=\frac{q\alpha}{\ln 2},\qquad
q=\ln 2\,f'(\alpha),\qquad
\alpha=\ln 2\,\tau'(q).
\eeq

In the present case,
exact numerical results demonstrate in an unambiguous way
that the scaling law~(\ref{Zsca}) holds for $q>0$, i.e., positive temperatures.
This observation corroborates and extends the findings of Mallows and Shepp~\cite{MS},
who have established rigorously that the scaling law~(\ref{Zsca})
holds whenever $q=2,3,\dots$ is a positive integer.

For negative temperatures, i.e., $q<0$,
the growth of the partition function is asymptotically governed
by the patterns whose probabilities are the smallest,
i.e., the two straight ones (see~(\ref{rising})).
The partition function therefore grows super-exponentially
as $Z_n(q)\approx2((n+1)!)^{\abs{q}}$, for any negative value of $q$.
This behavior leads to the breakdown of the multifractal formalism.
This is not an artifact which could be circumvented easily.
Indeed, there is actually a continuum of patterns,
including all the chirping ones,
which yield a logarithmically diverging effective rate $\alpha(n)$ (see~(\ref{acont})).
More generally, any quantity which has a high sensitivity
to the smallest of the probabilities~$P_n$
will be affected by logarithmic violations to scaling.

Let us proceed and describe quantitative results.

\begin{figure}[!ht]
\begin{center}
\includegraphics[angle=-90,width=.5\linewidth]{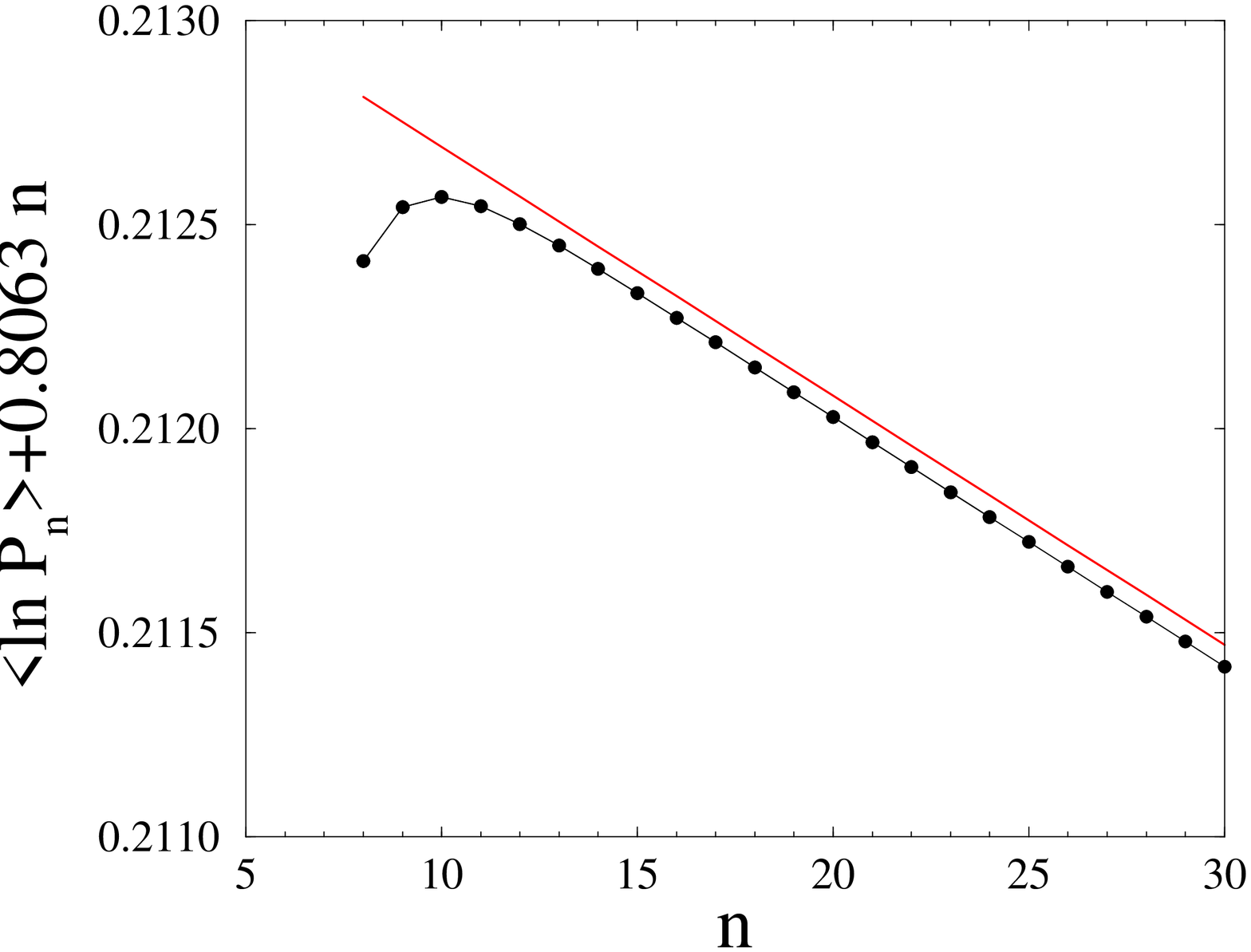}
\caption{\small
Plot of $\meaneps{\ln P_n}+0.8063 n$ against $n$ up to $n=30$.
The red line has slope $-61\times10^{-6}$.}
\label{a0}
\end{center}
\end{figure}

For $q=0$, we have $Z_n(0)=2^n$, hence $\tau(0)=-1$,
and so the support dimension takes the obvious value $D_0=1$.
More interestingly, taking the derivative of~(\ref{Zsca}) at $q=0$, we obtain
\beq
\meaneps{\ln P_n}=2^{-n}Z_n'(0)\approx-n\alpha_0.
\label{lnp}
\eeq
The very accurate numerical value
\beq
\alpha_0=\tau'(0)\ln 2=0.806361\dots
\label{alpha0}
\eeq
has been obtained by an exact numerical evaluation of $\meaneps{\ln P_n}$ up to $n=30$.
The very fast convergence of the data toward the asymptotic linear law~(\ref{lnp})
is illustrated in Figure~\ref{a0}.
A similar kind of convergence is observed for all subsequent quantities.

Considering higher-order derivatives of~(\ref{Zsca}) at $q=0$,
we conclude that all the cumulants of $\ln P_n$ are {\it extensive},
in the sense that they grow asymptotically linearly with $n$.
We have in particular
\beq
\vareps\ln P_n=\meaneps{(\ln P_n)^2}-\meaneps{\ln P_n}^2\approx nw_0,
\eeq
with
\beq
w_0=-\tau''(0)\ln 2=0.435600\dots
\eeq
As a consequence, and in physical terms,
the typical value $\alpha_0$ of the rate is self-averaging.
In other words, we have
\beq
\ln P_n\approx-n\alpha_0
\label{sav}
\eeq
for almost all (long enough) patterns.

Interestingly enough,
the typical rate $\alpha_0$ can also be interpreted as
the Lyapunov exponent of the random dynamical system
defined by the recursion~(\ref{prec}).
The {\it mean value} of the function $f_n(x)$,
obtained by averaging at each step the recursion~(\ref{prec}) over both values of $\eps_n$,
has the simple expression $\meaneps{f_n(x)}=2^{-n}$,
in agreement with the simple result $\meaneps{P_n}=2^{-n}$ (see~(\ref{pave})).
The {\it typical value} of the function $f_n(x)$
however keeps fluctuating in a non-trivial way,
and it falls off as $f_n(x)\sim\e^{-n\alpha_0}$,
i.e., exponentially faster than the mean value, as $\alpha_0-\ln 2=0.113214\dots>0$.

For $q=1$, as already mentioned,
the normalization of the probabilities $P_n$ ensures $Z_n(1)=1$.
More interestingly, taking the derivative of~(\ref{Zsca}) at $q=1$,
we predict that the entropy $\Sigma_n$ of the set $\{P_n\}$ grows as
\beq
\Sigma_n=-\sum_\epsn P_n(\epsn)\ln P_n(\epsn)\approx n\,D_1\,\ln 2,
\eeq
with
\beq
D_1=\tau'(1)=0.904475\dots
\eeq
This number is referred to as the entropy (information) dimension.

Higher integer values of the index $q$ are also of interest.
Indeed, as recalled above, the partition function $Z_n(q)$
is equal to the probability that $q$ independent random permutations
yield the same pattern.
We have $\tau(2)=0.856199\dots$ (hence $D_2=0.856199\dots$),
$\tau(3)=1.647144\dots$ (hence $D_3=0.823572\dots$), and so on.
The case of pairs of permutations ($q=2$)
has been investigated by analytical means by Mallows and Shepp~\cite{MS}.
These authors have determined the value of $D_2$ in terms of the smallest zero
of an explicit entire series.
Their approach however does not extend to higher values of $q$.

In the $q\to\infty$ limit,
the growth of the partition function is governed by the most probable patterns,
i.e., the alternating patterns, with rate $\alpha_\min=\ln(\pi/2)$ (see~(\ref{a+-})).
We thus get
\beq
D_\infty=\frac{\alpha_\min}{\ln 2}=\frac{\ln(\pi/2)}{\ln 2}=0.651496\dots
\eeq

\begin{figure}[!ht]
\begin{center}
\includegraphics[angle=-90,width=.485\linewidth]{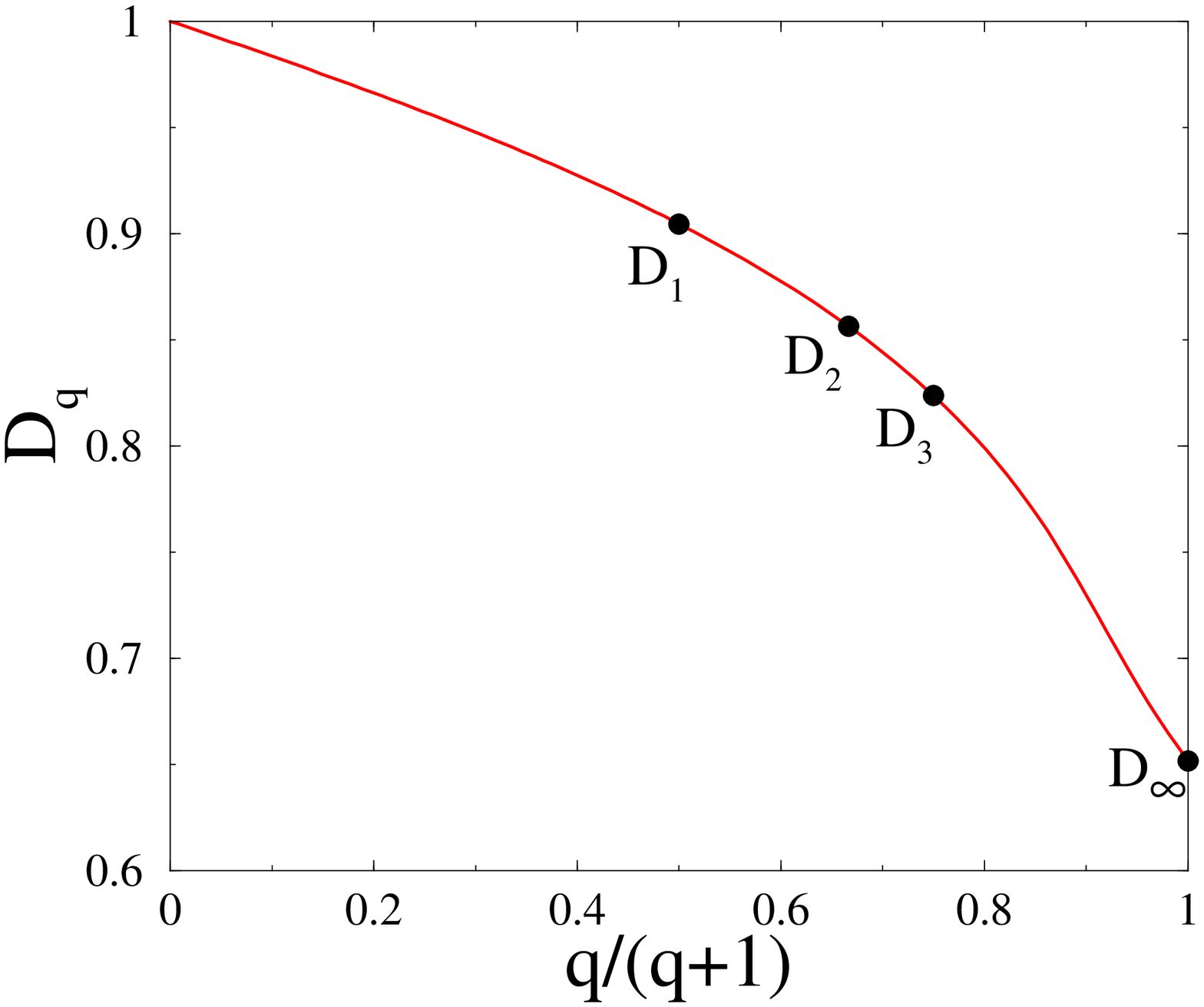}
\hfill
\includegraphics[angle=-90,width=.49\linewidth]{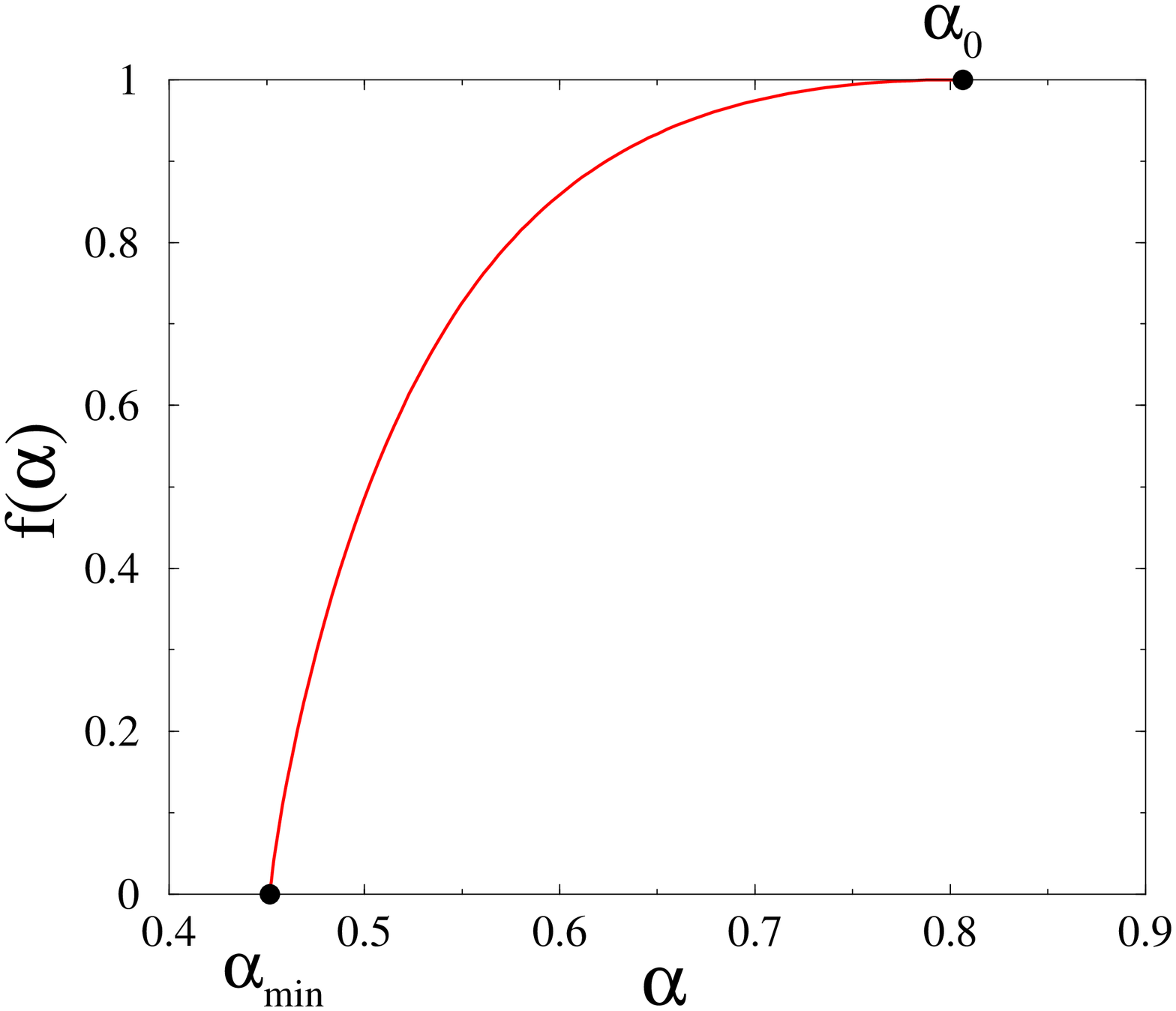}
\caption{\small
Left: generalized dimensions $D_q$ against $q/(q+1)$.
Right: multifractal spectrum $f(\alpha)$ against $\alpha$
in the range $\alpha_\min<\alpha<\alpha_0$.}
\label{dqfa}
\end{center}
\end{figure}

The main outcome of multifractal analysis is given in Figure~\ref{dqfa},
showing (left) the generalized dimensions $D_q$ (for $q>0$) against $q/(q+1)$
and (right) the multifractal spectrum $f(\alpha)$ against $\alpha$.
The latter only makes sense in the range $\alpha_\min\le\alpha\le\alpha_0$,
where $f(\alpha)$ grows from $f(\alpha_\min)=0$ to $f(\alpha_0)=1$.

\subsection{Patterns at fixed concentration of rises}

For uniformly chosen random patterns of length $n$,
we have seen that the logarithm of the probability $P_n$
is self-averaging and characterized by the typical rate $\alpha_0$.
The same self-averaging property holds for more general ensembles of random patterns.

A first interesting example consists in imposing the concentration~$c$ of rises,
i.e., in choosing at every place either a rise with probability $c$,
or a fall with the complementary probability:
\beq
\eps_n=\left\{\matrix{
+\hfill& \mbox{with probability}\hfill& c,\hfill\cr
-\hfill& \mbox{with probability}\hfill& 1-c.\hfill
}\right.
\label{ranc}
\eeq

\begin{figure}[!ht]
\begin{center}
\includegraphics[angle=-90,width=.5\linewidth]{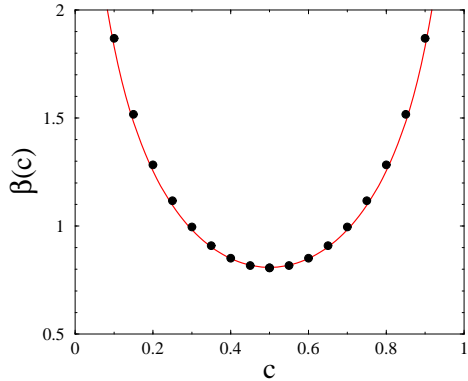}
\caption{\small
Effective typical rate $\beta(c)$ of the ensemble
of random patterns at fixed concentration $c$ of rises, against $c$.
Black symbols: numerical data.
Full red line: estimate~(\ref{lawc}).}
\label{beta}
\end{center}
\end{figure}

Within this ensemble, the logarithm of the probability $P_n$ is again self-averaging,
i.e., we have
\beq
\ln P_n\approx-n\beta(c),
\label{sac}
\eeq
where the effective typical rate $\beta(c)$ now depends on the concentration $c$ of rises.
Figure~\ref{beta} shows a plot of this quantity.
Each data point is obtained by averaging $\ln P_n$
over $10^5$ independent patterns of length $n=200$.

For $c=1/2$, the uniform ensemble is recovered,
and so the rate takes its minimal value $\beta(1/2)=\alpha_0=0.806361\dots$
As $c$ goes to 0 (resp.~1),
distances between consecutive rises (resp.~falls) become large.
More precisely, these distances are exponentially distributed,
with a mean value approximately equal to $1/c$ (resp.~$1/(1-c)$).
Following the line of thought which led us to~(\ref{periods}), we thus obtain
\beq
\beta(c)\approx-\ln(c(1-c))-\euler\quad(c\to\hbox{0 or 1}),
\label{lawc}
\eeq
where $\euler$ denotes Euler's constant.
This estimate provides a surprisingly good description
of the effective rate over the whole range of concentrations.

\subsection{Symmetric Markovian patterns}

Another interesting example consists of the patterns
where rises and falls are equally probable but correlated.
The null model for this case is the symmetric Markovian ensemble,
where $\eps_n$ is equal to $\eps_{n-1}$ with some persistence probability~$r$,
and to its opposite with the complementary probability:
\beq
\eps_n=\left\{\matrix{
+\eps_{n-1}\hfill& \mbox{with probability}\hfill& r,\hfill\cr
-\eps_{n-1}\hfill& \mbox{with probability}\hfill& 1-r.\hfill
}\right.
\label{ranm}
\eeq

\begin{figure}[!ht]
\begin{center}
\includegraphics[angle=-90,width=.48\linewidth]{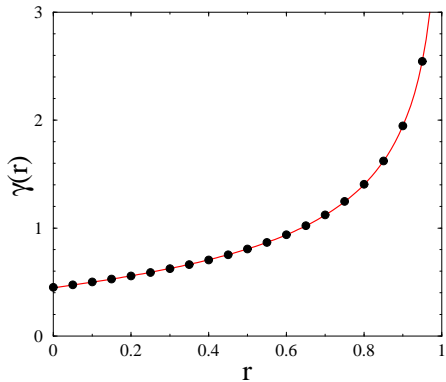}
\caption{\small
Effective typical rate $\gamma(r)$ of the symmetric Markovian ensemble of random patterns,
against the persistence probability $r$.
Black symbols: numerical data.
Full red line: two-parameter fit incorporating the estimate~(\ref{lawr}).}
\label{gamma}
\end{center}
\end{figure}

The logarithm of the probability $P_n$ is again self-averaging, i.e., we have
\beq
\ln P_n\approx-n\gamma(r),
\label{sar}
\eeq
where the effective typical rate $\gamma(r)$ depends on the persistence probability $r$.
Figure~\ref{gamma} shows a plot of this quantity.

For $r=1/2$, the uniform ensemble is again recovered,
and so we have $\gamma(1/2)=\alpha_0=0.806361\dots$
As $r$ goes to 0, a rise is followed by a fall with very high probability,
and vice versa.
As a consequence, a typical pattern of the ensemble
consists of long alternating stretches,
and so $\gamma(r)$ goes to $\alpha_\min=\ln(\pi/2)$ (see~(\ref{a+-})).
In the opposite limit ($r\to1$),
a typical pattern consists of long ordered stretches of rises and falls,
whose lengths are again exponentially distributed,
with a mean value scaling as $1/(1-r)$, and so
\beq
\gamma(r)\approx-\ln(1-r)-\euler\quad(r\to1),
\label{lawr}
\eeq
where $\euler$ again denotes Euler's constant.

\appendix

\section{Explicit correspondence between the probabilistic and combinatorial approaches}
\label{appa}

This appendix presents the explicit correspondence
between the probabilistic approach and the combinatorial one.
For a given pattern $\epsn$ of~$n$ rises and falls,
the probabilistic approach (see Section~\ref{proba})
involves the $n$th degree polynomial $f_n(x)$, which has $n+1$ coefficients,
while the combinatorial one (see Section~\ref{combi})
involves the $n+1$ integers $a_{n,j}$.
These two sets of numbers carry the same information.

The precise correspondence~(\ref{pa})
between both descriptions can be established as follows.
Consider a permutation $\s$ yielding the pattern $\epsn$ and such that $\s_n=j$.
The number $x_n$ is therefore the $(j+1)$st one
of the random numbers $x_0,\dots,x_n$ written in increasing order.
In other words, $f_{n,j}(x)\,\d x$ is the probability
of the following event:~$j$ numbers are in the interval $[0,\,x]$,
one number is in the interval $[x,\,x+\d x]$,
and the remaining $n-j$ numbers are in the interval $[x+\d x,\,1]$.
The latter probability is given by the multinomial formula
\beq
f_{n,j}(x)\,\d x=\frac{(n+1)!}{j!1!(n-j)!}\,x^j(\d x)^1(1-x-\d x)^{n-j},
\eeq
i.e.~(to first order in $\d x$),
\beq
f_{n,j}(x)=\frac{(n+1)!}{j!(n-j)!}\,x^j(1-x)^{n-j}.
\label{ftri}
\eeq

By definition, there are $a_{n,j}$ permutations $\s$ such that $\s_n=j$,
among a total of $(n+1)!$.
Summing the expression~(\ref{ftri}) over $j$ with the weights $a_{n,j}/(n+1)!$,
we get the result announced in~(\ref{pa}), i.e.,
\beq
f_n(x)=\sum_{j=0}^n a_{n,j}\,\frac{x^j(1-x)^{n-j}}{j!(n-j)!}.
\label{apppa}
\eeq
The inverse formula reads
\beq
a_{n,j}=j!(n-j)!\oint\frac{\d x}{2\pi\ii}\,\frac{f_n(x)}{x^{j+1}(1-x)^{n-j+1}},
\label{appap}
\eeq
where the integration contour circles once around the point $x=0$.

In order to be complete, let us check explicitly
that the recursions~(\ref{prec}) and~(\ref{arec}) are equivalent to each other.
Assume the probabilistic recursion~(\ref{prec}) holds,
and consider the case where $\eps_n=+$
(the other case can be dealt with in a similar way).
We have $f_n(0)=0$, implying $a_{n,0}=0$, and $f_{n-1}(x)=f_n'(x)$.
Now consider the difference $d_{n,j}=a_{n,j}-a_{n,j-1}$ for $j=1,\dots,n$.
The formula~(\ref{appap}) yields
\beqa
d_{n,j}
\!\!\!&=&\!\!\!
\oint\frac{\d x}{2\pi\ii}\,f_n(x)
\left(\frac{j!(n-j)!}{x^{j+1}(1-x)^{n-j+1}}-\frac{(j-1)!(n-j+i)!}{x^j(1-x)^{n-j+2}}\right)
\nonumber\\
\!\!\!&=&\!\!\!
\oint\frac{\d x}{2\pi\ii}\,f_n(x)
\left(-\frac{\d}{\d x}\,\frac{(j-1)!(n-j)!}{x^j(1-x)^{n-j+1}}\right)
\nonumber\\
\!\!\!&=&\!\!\!
\oint\frac{\d x}{2\pi\ii}\left(\frac{\d}{\d x}\,f_n(x)\right)
\frac{(j-1)!(n-j)!}{x^j(1-x)^{n-j+1}}\qquad(\hbox{by parts})
\nonumber\\
\!\!\!&=&\!\!\!
\oint\frac{\d x}{2\pi\ii}\,f_{n-1}(x)
\frac{(j-1)!(n-j)!}{x^j(1-x)^{n-j+1}}
\nonumber\\
\!\!\!&=&\!\!\!
a_{n-1,j-1}.
\eeqa
This completes the explicit check of the equivalence between
the probabilistic approach and the combinatorial one.

\section{Generalized hyperbolic and trigonometric functions}
\label{appb}

In this appendix we gather formulas and results
on generalized hyperbolic and trigonometric functions,
which are used in the study of periodic patterns
(sections~\ref{palt} and~\ref{parb}).
Both families of functions are sometimes~\cite{CS,CO}
referred to as Olivier functions~\cite{OLIV}.
Generalized hyperbolic functions are also described in~\cite{ghf}.

\subsubsection*{Generalized hyperbolic functions.}

These functions provide a useful basis of solutions to the differential equation
\beq
f^{(p)}=f,
\label{eqh}
\eeq
where $p\ge2$ is a given integer.
Looking for a solution of the form $\e^{ax}$,
we are left with the condition $a^p=1$.
We thus obtain a basis of $p$ exponential solutions:
\beq
A_j(x)=\e^{x\o^j}\quad(j=0,\dots,p-1),
\eeq
corresponding to $a=\o^j$, where
\beq
\o=\e^{2\pi\ii/p}
\label{odef}
\eeq
is the first $p$th root of unity.

It is advantageous to introduce the linear combinations
\beq
H_{p,q}(x)=\frac{1}{p}\sum_{j=0}^{p-1}\o^{-qj}\e^{x\o^j}\quad(q=0,\dots,p-1).
\label{hpq}
\eeq
The generalized hyperbolic functions thus defined
provide another basis of solutions to~(\ref{eqh}).
They obey the first-order differential equations
\beq
H_{p,q}'=H_{p,q-1}\quad(q=1,\dots,p-1),\qquad H_{p,0}'=H_{p,p-1}.
\eeq
The power-series expressions
\beq
H_{p,q}(x)=\sum_{k\ge0}\frac{x^{kp+q}}{(kp+q)!}
\label{hseries}
\eeq
are obtained by expanding the exponentials in~(\ref{hpq}).
Another advantage of these functions is the simple expressions
of their Laplace transforms:
\beq
\w H_{p,q}(s)=\frac{s^{p-q-1}}{s^p-1}.
\eeq
The first few generalized hyperbolic functions read
\beqa
H_{2,0}(x)
\!\!\!&=&\!\!\!
\cosh x,
\nonumber\\
H_{2,1}(x)
\!\!\!&=&\!\!\!
\sinh x,
\\
\nonumber\\
H_{3,0}(x)
\!\!\!&=&\!\!\!
\frac{1}{3}\left(\e^x+2\,\e^{-x/2}\cos\frac{x\sqrt{3}}{2}\right),
\nonumber\\
H_{3,1}(x)
\!\!\!&=&\!\!\!
\frac{1}{3}\left(\e^x-\e^{-x/2}\cos\frac{x\sqrt{3}}{2}+\sqrt{3}\,\e^{-x/2}\sin\frac{x\sqrt{3}}{2}\right),
\nonumber\\
H_{3,2}(x)
\!\!\!&=&\!\!\!
\frac{1}{3}\left(\e^x-\e^{-x/2}\cos\frac{x\sqrt{3}}{2}-\sqrt{3}\,\e^{-x/2}\sin\frac{x\sqrt{3}}{2}\right),
\\
\nonumber\\
H_{4,0}(x)
\!\!\!&=&\!\!\!
\frac{1}{2}(\cosh x+\cos x),
\nonumber\\
H_{4,1}(x)
\!\!\!&=&\!\!\!
\frac{1}{2}(\sinh x+\sin x),
\nonumber\\
H_{4,2}(x)
\!\!\!&=&\!\!\!
\frac{1}{2}(\cosh x-\cos x),
\nonumber\\
H_{4,3}(x)
\!\!\!&=&\!\!\!
\frac{1}{2}(\sinh x-\sin x).
\eeqa

\subsubsection*{Generalized trigonometric functions.}

The above construction can be transposed to the differential equation
\beq
f^{(p)}=-f.
\label{eqt}
\eeq
Looking again for a solution of the form $\e^{ax}$,
we are left with the condition $a^p=-1$.
We thus obtain a basis of $p$ exponential solutions:
\beq
B_j(x)=\e^{x\o^{j+1/2}}\quad(j=0,\dots,p-1),
\eeq
corresponding to $a=\o^{j+1/2}$.

We introduce the linear combinations
\beq
T_{p,q}(x)=\frac{1}{p}\sum_{j=0}^{p-1}\o^{-q(j+1/2)}\e^{x\o^{j+1/2}}\quad(q=0,\dots,p-1).
\label{tpq}
\eeq
The generalized trigonometric functions thus defined
provide another basis of solutions to~(\ref{eqt}).
They obey the first-order differential equations
\beq
T_{p,q}'=T_{p,q-1}\quad(q=1,\dots,p-1),\qquad T_{p,0}'=-T_{p,p-1}.
\eeq
The power-series expressions
\beq
T_{p,q}(x)=\sum_{k\ge0}(-1)^k\frac{x^{kp+q}}{(kp+q)!}
\label{tseries}
\eeq
are obtained by expanding the exponentials in~(\ref{tpq}).
Finally, their Laplace transforms are
\beq
\w T_{p,q}(s)=\frac{s^{p-q-1}}{s^p+1}.
\eeq
The first few generalized trigonometric functions read
\beqa
T_{2,0}(x)
\!\!\!&=&\!\!\!
\cos x,
\nonumber\\
T_{2,1}(x)
\!\!\!&=&\!\!\!
\sin x,
\\
\nonumber\\
T_{3,0}(x)
\!\!\!&=&\!\!\!
\frac{1}{3}\left(\e^{-x}+2\,\e^{x/2}\cos\frac{x\sqrt{3}}{2}\right),
\nonumber\\
T_{3,1}(x)
\!\!\!&=&\!\!\!
\frac{1}{3}\left(-\e^{-x}+\e^{x/2}\cos\frac{x\sqrt{3}}{2}+\sqrt{3}\,\e^{-x/2}\sin\frac{x\sqrt{3}}{2}\right),
\nonumber\\
T_{3,2}(x)
\!\!\!&=&\!\!\!
\frac{1}{3}\left(\e^{-x}-\e^{x/2}\cos\frac{x\sqrt{3}}{2}+\sqrt{3}\,\e^{-x/2}\sin\frac{x\sqrt{3}}{2}\right),
\label{t3}
\\
\nonumber\\
T_{4,0}(x)
\!\!\!&=&\!\!\!
\cosh\frac{x}{\sqrt{2}}\cos\frac{x}{\sqrt{2}},
\nonumber\\
T_{4,1}(x)
\!\!\!&=&\!\!\!
\frac{1}{\sqrt{2}}\left(\cosh\frac{x}{\sqrt{2}}\sin\frac{x}{\sqrt{2}}+\sinh\frac{x}{\sqrt{2}}\cos\frac{x}{\sqrt{2}}\right),
\nonumber\\
T_{4,2}(x)
\!\!\!&=&\!\!\!
\sinh\frac{x}{\sqrt{2}}\sin\frac{x}{\sqrt{2}},
\nonumber\\
T_{4,3}(x)
\!\!\!&=&\!\!\!
\frac{1}{\sqrt{2}}\left(\cosh\frac{x}{\sqrt{2}}\sin\frac{x}{\sqrt{2}}-\sinh\frac{x}{\sqrt{2}}\cos\frac{x}{\sqrt{2}}\right).
\label{t4}
\eeqa

\section*{Acknowledgments}

It is a pleasure to thank several participants to ALEA 2014,
and especially Nicolas Basset and Philippe Marchal, for very fruitful discussions.

\bibliography{mybibfile.bib}

\end{document}